\documentclass[journal,10pt,onecolumn,draftclsnofoot,]{IEEEtran}

\usepackage[USenglish]{babel}

\usepackage[T1]{fontenc}  
\usepackage{graphicx}
\usepackage[cmex10]{amsmath} 
\usepackage[tight,footnotesize]{subfigure} 

\usepackage{float} 
\usepackage{booktabs,colortbl} 

\usepackage{balance}

\usepackage[dvipsnames]{xcolor}

\usepackage{tabu}

\usepackage{eso-pic}

\usepackage{MMC_Notation}

\usepackage{multicol}

\usepackage{textcomp} 
\usepackage{gensymb} 

\usepackage{subfig}
\usepackage{bm} 

\usepackage{cite}

\usepackage{arydshln} 

\newcommand\scalemath[2]{\scalebox{#1}{\mbox{\ensuremath{\displaystyle #2}}}}

\newenvironment{bsmallmatrix}
{\left[\begin{smallmatrix}}
	{\end{smallmatrix}\right]}

\begin{document}
\title{Generalized Voltage-based State-Space Modelling of Modular Multilevel Converters with Constant Equilibrium in Steady-State}


\author{~Gilbert~Bergna-Diaz,
        Julian~Freytes,
        ~Xavier~Guillaud,~\IEEEmembership{Member,~IEEE,}
        \\~Salvatore~D'Arco,
        and~Jon~Are~Suul,~\IEEEmembership{Member,~IEEE}
\thanks{G. Bergna-Diaz is with the Norwegian University of Science and Technology (NTNU), Trondheim, Norway (gilbert.bergna@ntnu.no) \newline \indent J. Freytes and X. Guillaud are with the Universit\'{e} Lille, Centrale Lille, Arts et M\'{e}tiers, HEI - EA 2697,  L2EP -  Lille, France (julian.freytes@centralelille.fr,
xavier.guillaud@centralelille.fr) \newline \indent S. D'Arco and J. A. Suul are with SINTEF Energy Research, Trondheim, Norway, (salvatore.darco@sintef.no, jon.a.suul@sintef.no) }}

\maketitle
\IEEEpeerreviewmaketitle
\begin{abstract}
This paper demonstrates that the sum and difference of the upper and lower arm voltages are suitable variables for deriving a generalized state-space model of an MMC which settles at a constant equilibrium in steady-state operation, while including the internal voltage and current dynamics. The presented modelling approach allows for separating the multiple frequency components appearing within the MMC as a first step of the model derivation, to avoid variables containing multiple frequency components in steady-state. On this basis, it is shown that Park transformations at three different frequencies ($+\omega$, $-2\omega$ and $+3\omega$) can be applied for deriving a model formulation where all state-variables will settle at constant values in steady-state, corresponding to an equilibrium point of the model. The resulting model is accurately capturing the internal current and voltage dynamics of a three-phase MMC, independently from  how the control system is implemented. The main advantage of this model formulation is that it can be linearised, allowing for eigenvalue-based analysis of the MMC dynamics. Furthermore, the model can be utilized for control system design by multi-variable methods requiring any stable equilibrium to be defined by a fixed operating point. Time-domain simulations in comparison to an established average model of the MMC, as well as results from a detailed simulation model of an MMC with 400 sub-modules per arm, are presented as verification of the validity and accuracy of the developed model.
\end{abstract}


This manuscript is partly based on the following conference publication:
Gilbert Bergna, Jon Are Suul, Salvatore D'Arco, "State-Space Modeling of Modular Multilevel Converters for Constant Variables in Steady-State," presented at the 17th IEEE Workshop on Control and Modeling for Power Electronics, COMPEL 2016, Trondheim, Norway, 27-30 June 2016.
Compared to the initial conference paper, the text has been thoroughly revised, the presented model has been generalized and new results have been included. 

\section{Introduction}

\par The Modular Multilevel Converter (MMC) is emerging as the preferred topology for Voltage Source Converter (VSC) -based HVDC transmission schemes  \cite{Lesnicar2003,Adapa2012}. Especially in terms of its low losses, modularity, scalability and low harmonic content in the output ac voltage, the MMC topology provides significant advantages for HVDC applications compared to two- or three-level VSCs. However, the MMC is characterized by additional internal dynamics related to the circulating currents and the internal capacitor voltages of the upper and lower arms of each phase \cite{Antonopoulos2009,Harnefors2013}. Thus, the modelling, control and analysis of the MMC is more complicated than for other VSC topologies. 

\par
\begin{figure*}[tb]
	\centering %
	\includegraphics[width=1\columnwidth]{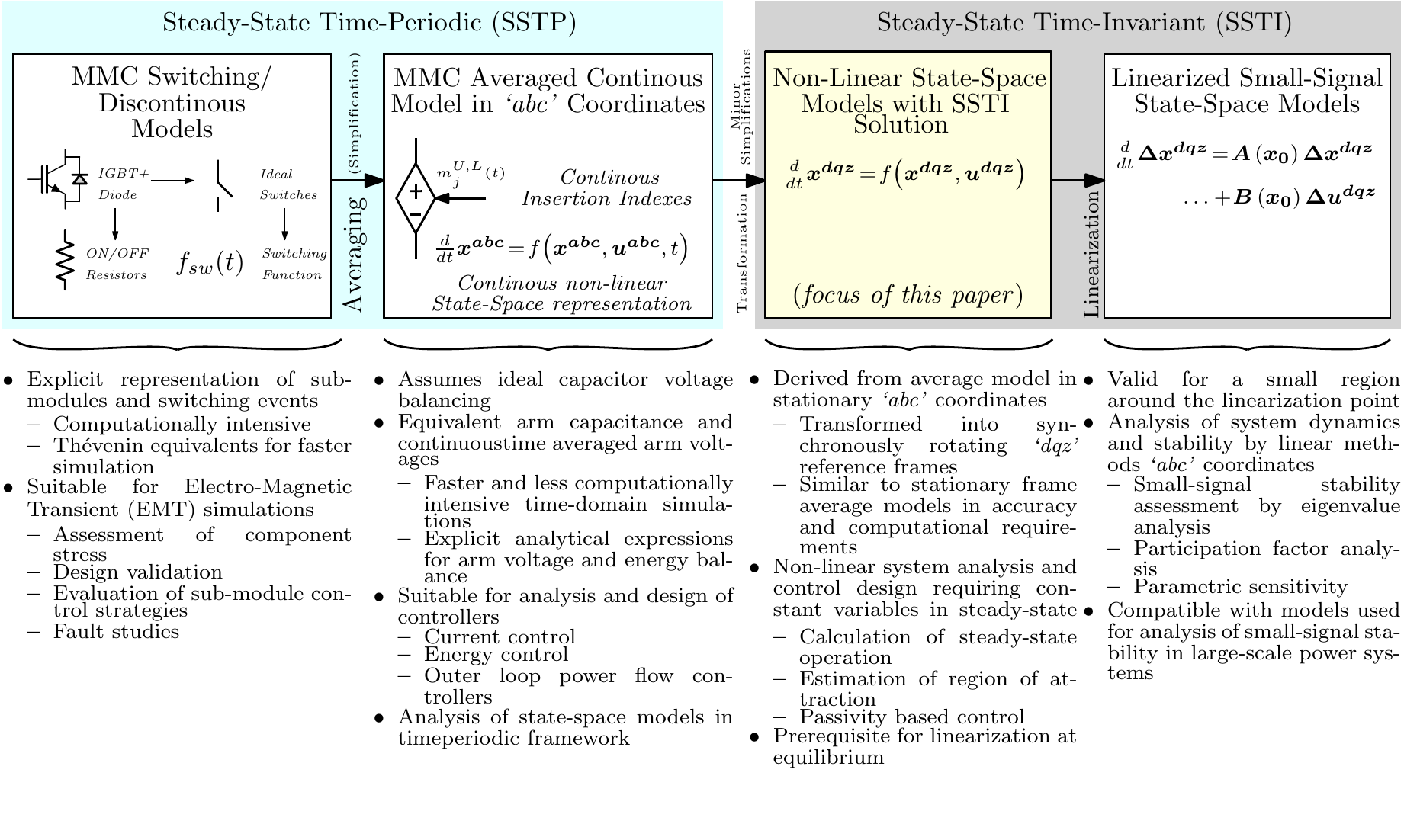}
  \caption{Overview of MMC modelling approaches and their areas of application}
	\label{Fig:StateSpace}
\end{figure*}

 Different types of studies are necessary for design and analysis of MMC-based HVDC transmission systems, requiring various detailing levels in the modelling. A general overview of MMC modelling approaches suitable for different types of studies is shown in Fig. \ref{Fig:StateSpace}. The most detailed models allow for simulating the switching operations of the individual sub-modules of the MMC, as shown to the right of the figure. Such models can be used for studying all modes of operation and all the control loops of the MMC, including the algorithms for balancing the sub-module voltages. If equal voltage distribution among the sub-modules in each arm of an MMC can be assumed, average arm models (AAM) can be introduced. The AAM modelling approach allows for representing each arm of the MMC by a controllable voltage source associated with a corresponding equivalent capacitance, and is introducing a significant reduction of complexity while still maintaining an accurate representation of the internal dynamics.  \cite{Antonopoulos2009,Harnefors2013,Rohner2011,Delarue2013}. 

\par Average modelling by the AAM representation, or by equivalent energy-based models, are suitable for simplified simulations and analysis, and have been widely used as basis for control system design \cite{Harnefors2013, Delarue2013,Saad2015}. However, the variables of such models are Steady State Time Periodic (SSTP), with the currents and capacitor voltages in each arm of the MMC containing multiple frequency components \cite{Ilves2012}. This prevents a straightforward application of the Park transformation for obtaining state-space models of three-phase MMCs represented in a single Synchronously Rotating Reference Frame (SRRF), according to the modelling approaches commonly applied for control system design and small-signal stability analysis of two-level VSCs \cite{Chaudhuri2011,Uhlen2012, Beerten2016}. However, for obtaining a linearized small-signal model of an MMC that can be analyzed by traditional techniques for eigenvalue-based stability analysis, it is necessary to derive a SRRF state-space model with a Steady-State Time Invariant (SSTI) solution. As indicated in Fig. 1, such a SRRF $dqz$ model must be derived from an equivalent average model in the stationary $abc$ coordinates. If a non-linear model with a SSTI solution, corresponding to defined equilibrium point, can be obtained, a Linear Time Invariant (LTI) model suitable for eigenvalue analysis can be directly obtained by linearization.

\par Several approaches for obtaining LTI state-space models of MMCs have been recently proposed in the literature, motivated by the need for representing MMC HVDC transmission systems in eigenvalue-based small-signal stability studies. The simplest approach has been to neglect parts of the internal dynamics of the MMC, and model mainly the ac-side dynamics in a SRRF together with a simplified dc-side representation, as in the models proposed in \cite{Liu2014,Ludois2014,Trinh2016}. However, if the dynamics associated with the internal equivalent capacitor voltages of the MMC and the interaction with the circulating currents are ignored, such models will imply significant inaccuracies. Especially if a power balance between the ac- and dc-sides of the converter is assumed in the same way as for a two-level VSC model, like in \cite{Liu2014,Trinh2016}, the model will only be suitable for representing very slow transients. Therefore, more detailed dynamic state-space models have been proposed in \cite{BergnaECCE2015,Freytes2016,Far2015,Jamshidifar2016,Najmi2015,BergnaCOMPEL2016,GolePhasors2016}. These available models have been developed for representing two different cases:
\begin{enumerate} 
\item 
The approaches presented in \cite{BergnaECCE2015,Freytes2016} are based on the assumption that the modulation indices for the MMC arms are calculated to compensate for the voltage oscillations in the internal equivalent arm capacitor voltages, referred to as Compensated Modulation (CM). This strategy for control system implementation limits the coupling between the internal variables of the MMC and the ac- and dc-side variables. Thus, CM-based control allows for simplified modelling of the MMC, where only the aggregated dynamics of the zero-sequence circulating current and the total energy stored in the capacitors of the MMC are represented. As a result, these models can provide accurate representation of the ac- and dc-side terminal behavior of MMCs, but imply that the dynamics of the internal variables cannot be analyzed.
\item
The approaches proposed in \cite{Far2015,Jamshidifar2016,Najmi2015,GolePhasors2016} consider all the internal variables of the MMC, under the assumption of a control system with a Circulating Current Suppression Controller (CCSC) implemented in a negative sequence double frequency SRRF  \cite{Tu2012}. Indeed, the methods proposed in \cite{Far2015,Jamshidifar2016,GolePhasors2016} model the MMC by representing the internal second harmonic circulating currents and the corresponding second harmonic arm voltage components in a SRRF rotating at twice the fundamental frequency. However, the harmonic superposition principles assumed in the modelling, corresponding to phasor-based representation, could affect the information about the non-linear characteristics of the MMC, and correspondingly limit the applicability of the models in non-linear techniques for analysis and control system design. A similar approximation was also made when separately modelling the fundamental frequency and the second harmonic frequency dynamics of the upper and lower arm capacitor voltages in \cite{Najmi2015}. 
\end{enumerate}

The main contribution of this paper is to present a linearizable SSTI state-space representation of an MMC with as few simplifications as possible in the derivation of the model. Indeed, the presented approach is intended for preserving the fundamental non-linearity of the stationary frame average model of the MMC that is used as starting point for the presented derivations. This is achieved by utilizing the information about how the different variables of the MMC contain mainly combinations of dc-components, fundamental frequency components and double frequency oscillations in steady state operation. By using the sum ($\Sigma$) and difference ($\Delta$) between the variables of the upper and lower arms of the MMC as variables, a natural frequency separation can be obtained where the $\Delta$ variables contain only a fundamental frequency component while $\Sigma$ variables contain dc and 2$\omega$ components. This frequency separation allows for applying appropriate Park transformations to each set of variables, resulting in an SSRF model where all state variables settle to a constant equilibrium point in steady-state operation. 
Thus, the obtained model is suitable for non-linear control system design, for instance by applying passivity theory \cite{Marcelo,PassivePI}, but can also be directly linearized to obtain a detailed small-signal model representing the dynamic characteristics of an MMC.

The first contribution to this modelling approach was presented in \cite{BergnaCOMPEL2016}, but this paper will extends the derivations from \cite{BergnaCOMPEL2016} to obtain a model that is applicable independently from the applied approach for calculating the modulation indices of the MMC. Furthermore, the model derivation has been expanded to include the effect of the zero-sequence of the difference between upper and lower insertion indexes $m_z$ in the MMC dynamics, which was neglected in \cite{BergnaCOMPEL2016}. This extension of the model can be useful when third harmonic injection is used for increasing the voltage utilization \cite{Houldsworth1984,Norrga2012}, and in case a zero sequence component in the output voltage is utilized to control the energy distribution within the MMC. 

The applied modelling approach and the derivations required for obtaining the presented generalized voltage-based state-space model of an MMC with SSTI characteristics are presented in detail, since similar techniques can also be useful for modelling and analysis of MMC control strategies implemented in the stationary frame. The validity of the derived model is demonstrated by time-domain simulations in comparison to the average model used as starting point for the derivations, and the validity of the obtained results are confirmed in comparison to a detailed simulation model of an MMC with 400 sub-modules per arm.

\section{MMC Modelling in the stationary reference frame: Topology, $\Sigma$-$\Delta$ vector representation and frequency analysis}\label{Sec:MMC_Model_and_control}
\subsection{Average model representation of the MMC topology}

\par
\begin{figure}[t!]
	\centering
	\includegraphics[width=0.5\columnwidth]{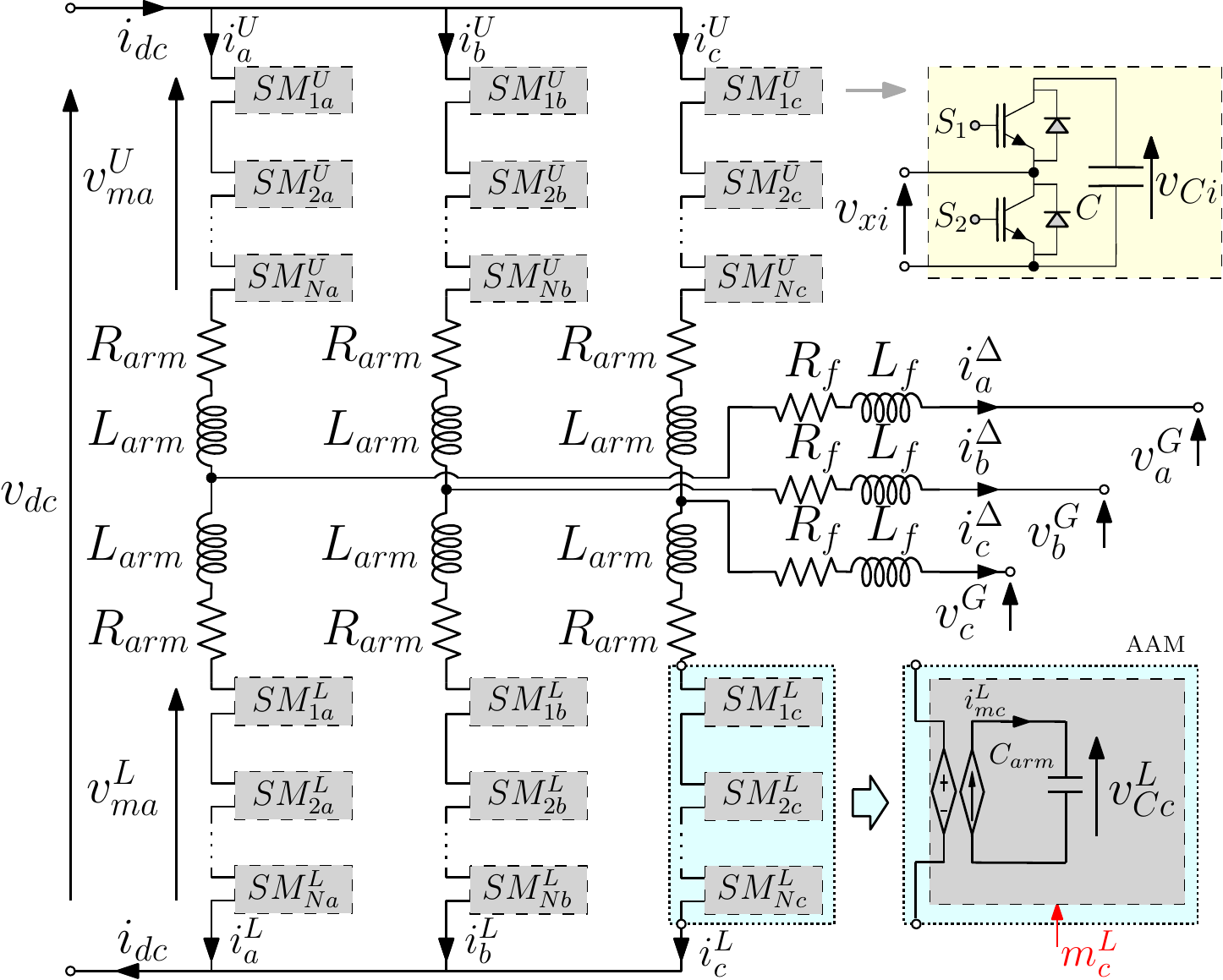}
  \caption{MMC Topology and AAM for the lower arm (phase $C$)}
	\label{Fig:MMC_Topology_and_AAM_2}
\end{figure}

 The basic topology of a three-phase MMC is synthesized by the series connection of  $N$ sub-modules (SMs) with independent capacitors $C$ to constitute one arm of the converter as indicated by Fig.~\ref{Fig:MMC_Topology_and_AAM_2}. The sub-modules in one arm are connected to a filter inductor with equivalent inductance $L_{arm}$ and resistance $R_{arm}$ to form the connection between the dc terminal and the ac-side output. Two identical arms are connected to the upper and lower dc-terminals respectively to form one leg for each phase $j$ ($j \!=\!a,b,c$). The AC side is modeled with an equivalent resistance and inductance $R_f$ and $L_f$ respectively \cite{Saad2014}.

\par Assuming that all the SMs capacitors voltages are maintained in a close range, the series connection of submodules in each arm can be replaced by a circuit-based average model, corresponding to the well-known Arm Averaged Model (AAM) as indicated in Fig.~\ref{Fig:MMC_Topology_and_AAM_2} for the lower arm of phase $c$ \cite{Harnefors2013,Saad2015}. If the MMC is modelled by the AAM representation, each arm is appearing as a controlled voltage source in the three-phase topology, while a power balance is established between the arm and its equivalent capacitance. Thus, each arm can be represented by a conventional power-balance-based average model of a VSC, with a modulated voltage source interfacing the filter inductor, and a current source interfacing the capacitor-side. 
\par The output of the controlled voltage and current sources of the AAM, are here referred as the modulated voltages $v^U_{mj}$ and $v^L_{mj}$ and modulated currents $i^U_{mj}$ and $i^L_{mj}$, for the upper ($U$) and lower ($L$) arms of a generic phase $j$, and are described by the following equations: 

\begin{equation}
\begin{array}{cccc}
\vMUj= \mUj \vCUj,\qquad &\vMLj= \mLj \vCLj,\qquad   & i^U_{mj} = m^U_j i^U_j, \qquad &i^L_{mj} = m^L_j i^L_j   
\end{array} \label{Eq:ModulatedVariables}
\end{equation}

\noindent where $v^U_{Cj}$ and $v^L_{Cj}$ are respectively the voltages across the upper and lower arm equivalent capacitors; $m^U_j$ and $m^L_j$ are  the corresponding insertion indexes 
for the upper and lower arms, and $i^U_{j}$ and $i^L_{j}$ are the currents in the upper and lower arms, respectively.

\subsection{Modelling of the MMC with {$\Sigma$-$\Delta$} variables in the stationary \emph{abc} frame}
\par As mentioned in the introduction, the proposed approach adopts the $\Sigma$-$\Delta$ representation as opposed to the more common one based on the \textit{Upper-Lower} ($U$-$L$) arm notation, to ease the derivation of an MMC model with SSTI solution. More precisely, under this $\Sigma$-$\Delta$ representation, it is possible to initially classify the 11 states and 6 control variables for an average model of a three-phase MMC into two frequency groups; i.e., the $\Delta$ variables which are associated to the fundamental frequency $\omega$, and the $\Sigma$ variables which are in turn associated to $-2\omega$, and will be further discussed in section \ref{sec:freq}.
 It is therefore useful to redefine the voltages and currents that are defined in Fig. \ref{Fig:MMC_Topology_and_AAM_2} using this nomenclature, resulting in \eqref{Eq:iDSvDS_def}. Indeed, $\iDeltaj$ is the current flowing through the AC-side grid, whereas $\iSigmaj$ is the well-known circulating current of the MMC. Moreover, $\vCDeltaj$ and $\vCSigmaj$ are respectively the difference and the sum of voltages across the upper and lower equivalent capacitors.
\begin{equation}
\iDeltaj \defeq \IarmUj - \IarmLj,\qquad \iSigmaj \defeq {\left(\IarmUj + \IarmLj\right)}/{2},\qquad \vCDeltaj \defeq (\vCUj - \vCLj)/{2}, \qquad \vCSigmaj \defeq (\vCUj + \vCLj)/{2},
\label{Eq:iDSvDS_def}
\end{equation}
 
\par In addition, it is also useful to define the modulated voltages  given in \eqref{Eq:ModulatedVariables} in the $\Sigma$-$\Delta$ representation as in \eqref{eq:vmDS}, as well as modulation indexes as in \eqref{eq:mDS}.
\begin{equation}
\vMDeltaj \defeq \frac{- \vMUj + \vMLj}{2},\hspace{8mm}
\vMSigmaj \defeq \frac{\vMUj + \vMLj}{2} \label{eq:vmDS}
\end{equation}
\begin{equation}
\mDeltaj \defeq \mUj - \mLj,\hspace{8mm}
\mSigmaj \defeq \mUj + \mLj\label{eq:mDS} 
\end{equation}
 
\subsubsection{AC-grid current dynamics}

\par The three-phase ac-grid currents dynamics $\iDeltaabc$ are expressed using vector nomenclature in the stationary frame as in  \eqref{Eq:iDeltaABC}, 
\begin{equation}\label{Eq:iDeltaABC}
  \Leqac{\frac{d\iDeltaabc}{dt}} = \vMDeltaabc - \Vgabc  - \Reqac\iDeltaabc,
\end{equation}

\noindent where $\Vgabc$ is the grid voltage vector defined as $[\Vga \; \Vgb \; \Vgc]^\top$, whereas $\vMDeltaabc$ is the modulated voltage driving the ac-grid current defined as $[\vMDeltaa \; \vMDeltab \; \vMDeltac]^\top$, or more precisely as:
\begin{equation}
\vMDeltaabc = - \frac{1}{2}\left( \mDeltaabc \LCdot \vCSigmaabc + \mSigmaabc \LCdot \vCDeltaabc \right)
 \label{Eq:vMDeltaabc}
\end{equation}

\noindent where the \textit{upper} and \textit{lower} modulation indexes and voltage variables were replaced by their $\Sigma$-$\Delta$ equivalents for convenience. It is worth noticing that the operator ``$\LCdot$'' will be used here to represent the element-wise multiplication of vectors (e.g. $\begin{bsmallmatrix}a \\b \end{bsmallmatrix}\LCdot\begin{bsmallmatrix}c \\d \end{bsmallmatrix}=\begin{bsmallmatrix}ac \\bd \end{bsmallmatrix}$). Furthermore, $\Reqac$ and $\Leqac$ are the equivalent ac resistance and inductance, respectively defined as $\Rf +{\Rarm}/2$ and $\Lf +{\Larm}/2$.
 
\subsubsection{Circulating current dynamics}

\par The three-phase circulating currents dynamics in the stationary frame can be written by using vector notation as:
\begin{equation}\label{Eq:iSigmaABC}
 \Larm \frac{d\iSigmaabc}{dt} =  \frac{\bm{v_{dc}}}{2} - \vMSigmaabc - \Rarm \iSigmaabc,
\end{equation}

\noindent where $\bm{v_{dc}}$ is defined as $[\Vdc \; \Vdc \; \Vdc]^\top$ and $\vMSigmaabc$ is the modulated voltage driving the circulating current defined as $[\vMSigmaa \; \vMSigmab \; \vMSigmac]^\top$, or more precisely as:
\begin{equation}
\vMSigmaabc = \frac{1}{2}\left( \mSigmaabc \LCdot \vCSigmaabc + \mDeltaabc \LCdot \vCDeltaabc \right), \label{Eq:vMSigmaabc}
\end{equation} 

\noindent where the \textit{upper} and \textit{lower} modulation indexes and voltage variables were replaced by their $\Sigma$-$\Delta$ equivalents for convenience here as well.

\subsubsection{Arm capacitor voltage dynamics}

\par Similarly, the dynamics of the voltage sum and difference between the equivalent capacitors of the AAM can be expressed respectively as in \eqref{Eq:vCSigmaABCdef} and \eqref{Eq:vCDeltaABCdef}.
\begin{align}
2\Carm \frac{d \vCSigmaabc}{dt}   = \mDeltaabc \LCdot \frac{\iDeltaabc}{2}  + \mSigmaabc \LCdot \iSigmaabc \label{Eq:vCSigmaABCdef} \\
2\Carm \frac{d\vCDeltaabc}{dt} = \mSigmaabc \LCdot \frac{\iDeltaabc}{2}  + \mDeltaabc  \LCdot \iSigmaabc \label{Eq:vCDeltaABCdef}
\end{align}
 
\subsection{Simplified frequency analysis of the $\Sigma$-$\Delta$ variables}
\label{sec:freq}

\par It is well known that under normal operating conditions the grid current of the MMC  $\iDeltaabc$ oscillates at the grid frequency $\omega$, whereas the circulating current consists of a dc value or a dc value in addition to oscillating signals at $-2\omega$, depending on whether the second harmonic component is eliminated by control or not. Therefore, the following analysis will only focus on the remaining voltage states $\vCSigmaabc$ and $\vCDeltaabc$.
\par A simplified analysis can be performed by assuming that $\mUj$ is phase-shifted approximately $180\degree$ with respect to $\mLj$, resulting in $\mSigmaj \approx 1$ and $\mDeltaj \approx \hat{m}cos\left(\omega t\right)$. By inspecting the right-side of \eqref{Eq:vCSigmaABCdef}, it can be seen that in steady-state, the first product $\mDeltaj \iDeltaj/2$ gives a dc value in addition to an oscilatory signal at $-2\omega$, while the second product $\mSigmaj \iSigmaj$ gives a dc value in case a constant value of $\iSigmaabc$ is imposed by control (e.g. by CCSC \cite{Tu2012}), or a $2\omega$ signal otherwise, resulting for both cases in $2\omega$ oscillations in $\vCSigmaj$. 

\par Similarly for $\vCDeltaj$, the first product on the right-side of \eqref{Eq:vCDeltaABCdef}, $\mSigmaj \iDeltaj/2$, oscillates at $\omega$, while the second product $\mDeltaj \iSigmaj$ oscillates at $\omega$ in the case the CCSC is used or will result in a signal oscillating at $\omega$ superimposed to one at $3\omega$ otherwise. Note that if the assumption $\mSigmaj \approx 1$ is no longer considered, but instead $\mSigmaabc$ is allowed to have a second harmonic component superimposed to its dc value, the first term of \eqref{Eq:vCDeltaABCdef} will also produce an additional component at $3\omega$.
\par As will be shown in the remainder of the paper, this additional 3rd harmonic in the $\Delta$ variable does not significantly affect the initial frequency classification of the variables as it will be captured and isolated by the zero-sequence component after the application of Park's transformation at $\omega$ without affecting its corresponding $dq$ components. This is similar to the case for the $\Sigma$ variables, as in addition to the $-2\omega$ signals, they present a dc value. Here too, this additional (dc) value is isolated by the zero-sequence after the application of Park's transformation at $-2\omega$, without  affecting its $dq$ components. 
\par This initial classification of the state and control variables according to their main oscilatory frequency is summarized in Table~\ref{Table:Omegavs2Omega} and is considered the base for the methodology which is presented in the next section. 

\begin{table}[hbt]
\renewcommand{\arraystretch}{1.3}
\caption{MMC variables in $\Sigma$-$\Delta$ representation}
\label{Table:Omegavs2Omega}
\centering
    \begin{tabular}{c|c}
    \toprule
    Variables oscillating at $\omega$  & Variables oscillating at $-2\omega$      \\
    \midrule
    $\iDeltaj     = \IarmUj - \IarmLj$      &  $\iSigmaj   = (\IarmUj + \IarmLj)/{2}$  \\
    $\vMDeltaj = (- \vMUj + \vMLj)/{2}$   &  $\vMSigmaj = (\vMUj + \vMLj)/{2}$      \\
    $\mDeltaj = \mUj - \mLj$            &  $\mSigmaj  = \mUj + \mLj$        \\
    \bottomrule
    \end{tabular}
\end{table}

\section{Non-linear time-invariant MMC model with {$\Sigma$-$\Delta$} representation in \emph{dqz} frame and voltage-based formulation}\label{Sec:Chapter_MMC_dqz}

\par In this section, the derivations needed for obtaining the state-space time-invariant representation of the MMC with voltage-based formulation is presented in detail on basis of the approach from \cite{BergnaCOMPEL2016}. 


\par
\begin{figure}[tb]
	\centering
	\includegraphics[width=0.6\columnwidth]{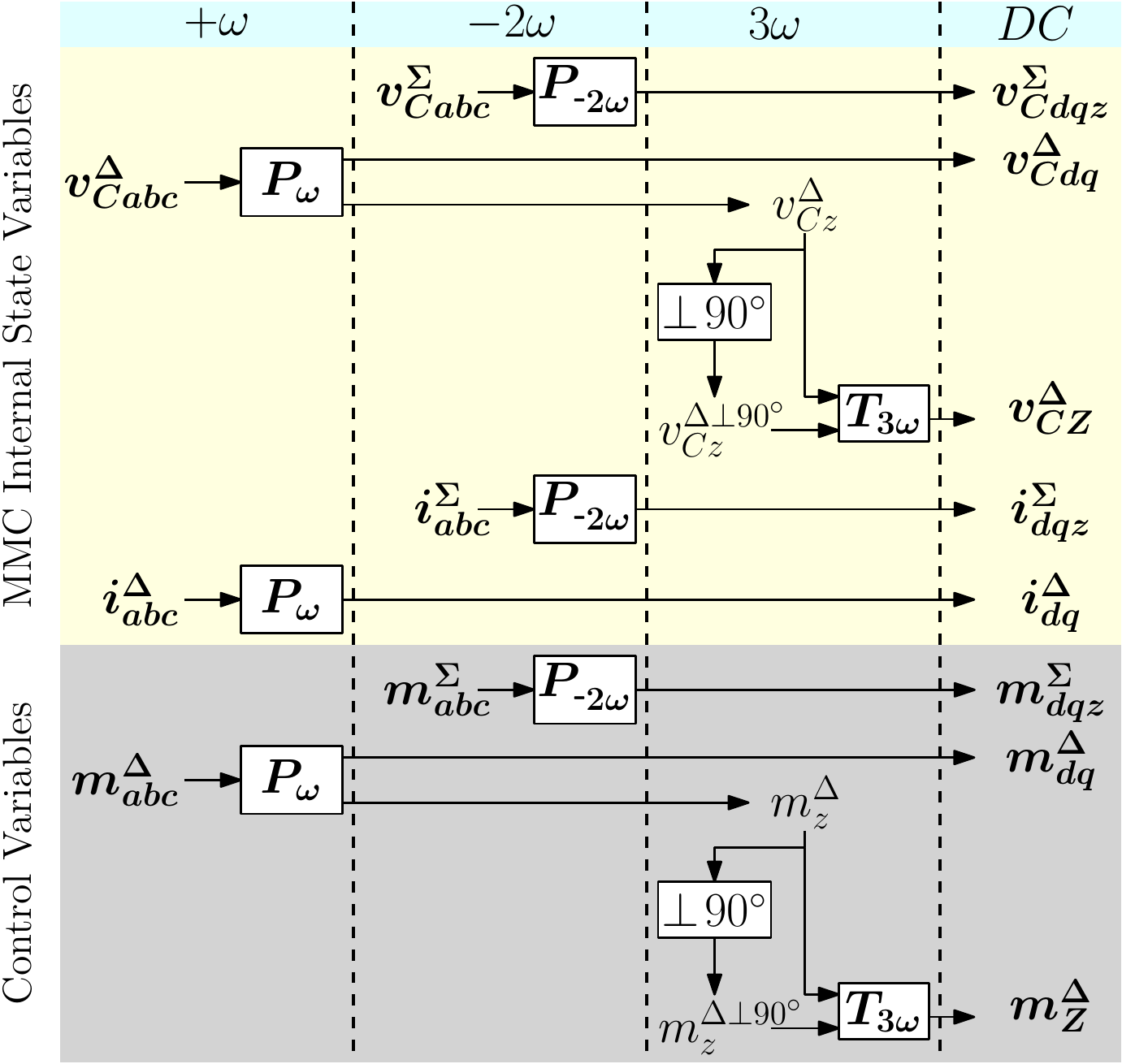}
  \caption{The proposed modelling approach based on three Park transformations to achieve SSTI control and state variables}
	\label{Fig:SSModelOverview}
\end{figure}

 The formulation of the MMC variables such that the initial separation of frequency components can be achieved constitutes the basis for the proposed modelling approach, as illustrated in Fig. \ref{Fig:SSModelOverview}. This figure indicates that Park transformations at different frequencies will be used to derive dynamic equations for equivalent $dqz$ variables that are SSTI in their respective reference frames. More precisely, the $\Delta$-variables ($\vCDeltaabc$, $\iDeltaabc$ and $\mDeltaabc$) are transformed into their $dqz$ equivalents by means of a Park transformation $\Parkw$ at the grid fundamental frequency $\omega$. By contrast, the $\Sigma$-variables ($\vCSigmaabc$, $\iSigmaabc$ and $\mSigmaabc$) are transformed into their $dqz$ equivalents by means of a Park transformation $\ParkTw$ at twice the grid frequency in negative sequence, $-2\omega$. In addition, a Park transformations at $3\omega$ will be used to ensure a SSTI representation of the zero sequence of the voltage difference $\vCDeltaz$, as well as for the zero sequence of the modulation index difference $\mDeltaz$.

\par In the remainder of this section, the mathematical derivation of dynamic equations with SSTI solution representing the dynamics of a three phase MMC will be expressed by using the approach illustrated by Fig. \ref{Fig:SSModelOverview}. The mathematical reformulation consists in expressing the vector variables in the stationary $abc$ frame as a function of their $dqz$ equivalents at their respective rotating frequencies.

\subsection{Voltage difference SSTI dynamics derivation}   
\subsubsection{Initial formulation}

\par The SSTI dynamics for the voltage difference is derived in the following. The starting point is indeed the SSTP dynamics of the variable given in \eqref{Eq:vCDeltaABCdef}, and recalled in \eqref{Eq:vCDeltaABC} for convenience. The first step consists in expressing the $abc$ vectors in the stationary frame as functions of their respective $dqz$ equivalents. This can be seen in the second line of \eqref{Eq:vCDeltaABC}, where $\vCDeltaabc$, $\mSigmaabc$, $\iDeltaabc$, $\mDeltaabc$ and $\iSigmaabc$ have been respectively replaced by $\ParkInvw \vCDeltadqz$, $\ParkInvTw \mSigmadqz$, $\ParkInvw \iDeltadqz$, $\ParkInvw \mDeltadqz$ and $\ParkInvTw \iSigmadqz$. Notice that the choice of using the inverse Park transformation matrix at $\omega$ ($\ParkInvw$) or at $2\omega$ ($\ParkInvTw$) is according to the frequency separation of the variables given in Table \ref{Table:Omegavs2Omega} and Fig. \ref{Fig:SSModelOverview}.
\begin{gather}
2\Carm \frac{d\vCDeltaabc}{dt} = \mSigmaabc \LCdot \frac{\iDeltaabc}{2}  + \mDeltaabc  \LCdot \iSigmaabc =\dots \nonumber \\
\underbrace{2\Carm   \frac{d  \ParkInvw}{dt} \vCDeltadqz  +  2\Carm \ParkInvw\frac{d  \vCDeltadqz}{dt}}_{\bm{\Phi_A^{\Delta}}} = \underbrace{\ParkInvTw \mSigmadqz \LCdot \frac{\ParkInvw \iDeltadqz}{2}}_{\bm{\Phi_B^{\Delta}}} 
+ \underbrace{\ParkInvw \mDeltadqz \LCdot \ParkInvTw \iSigmadqz}_{\bm{\Phi_C^{\Delta}}} \label{Eq:vCDeltaABC}  
\end{gather}

\par The equation expressed in \eqref{Eq:vCDeltaABC}, must be multiplied by the Park transformation matrix at the angular frequency $\omega$, so that it can be possible to solve for $d \vCDeltadqz /dt$. 

\par Multiplying ${\bm{\Phi_A^{\Delta}}}$ by $\Parkw$, gives:
\begin{gather}
\Parkw {\bm{\Phi_A^{\Delta}}}  =  {2\Carm   \Jw \vCDeltadqz  +  2\Carm  \frac{d  \vCDeltadqz}{dt}}\label{Eq:vCDeltaPartA}
\end{gather}

\noindent where $\Jw$ is defined as in \eqref{Eq:Jcoupledw}:
\begin{equation}\label{Eq:Jcoupledw}
\Jw  \defeq  
\begin{bmatrix}
  {0}&{\omega}&{0} \\ 
  { -\omega}&{0}&{0} \\ 
  {0}&{0}&{0} 
\end{bmatrix}
\end{equation}

\noindent Furthermore, multiplying ${\bm{\Phi_B^{\Delta}}}$ by $\Parkw$ gives:
\begin{gather} \label{Eq:vCDeltaPartB}
  	\Parkw {\bm{\Phi_B^{\Delta}}}  =   \Parkw \left(\ParkInvTw \mSigmadqz \LCdot \frac{\ParkInvw \iDeltadqz}{2} \right) =  \bm{M_{\Phi_B}^{\Delta}} 
  {\begin{bmatrix} 
  \iDeltad & 
  \iDeltaq &
  \color{gray}\iDeltaz   
\end{bmatrix}^\top}   
 \end{gather}

\noindent where $\bm{M_{\Phi_B}^{\Delta}}$ is expressed in \eqref{Eq:M_phiB_delta}. For simplicity, it will be considered that the system under study does not allow for the existence of the zero-sequence grid current; i.e., $\iDeltaz = 0$, highlighted in gray in \eqref{Eq:vCDeltaPartB}. Under this assumption, only the $dq$ component of \eqref{Eq:vCDeltaPartB} is time-invariant, as the $3\omega$ oscillatory signals that appear in $\bm{M_{\Phi_B}^{\Delta}}$ are either multiplying $\iDeltaz$ (third column of the matrix) or appear in the last row. However, it is possible to rewrite also the dynamics of $\vCDeltaz$ in SSTI form by means of additional mathematical manipulations, as will be shown further.
\begin{gather} \label{Eq:M_phiB_delta}
\bm{M_{\Phi_B}^{\Delta}}  =  
\frac{1}{4} 
\left[
\begin{array}{cc:c}
{\mSigmad + 2\mSigmaz}   & {-\mSigmaq} & \color{gray}{\mSigmad \cos(3\omega t) - \mSigmaq \sin(3\omega t)} \\ 
{- \mSigmaq}  & {- \mSigmad + 2\mSigmaz }& \color{gray}{\mSigmaq \cos(3\omega t) + \mSigmad \sin(3\omega t)} \\ \hdashline
{\mSigmad \cos(3\omega t) - \mSigmaq \sin(3\omega t)}   & {\mSigmaq \cos(3\omega t) + \mSigmad \sin(3\omega t)}  & \color{gray}{2\mSigmaz} 
 \end{array}
 \right]
 \end{gather}
\par Finally, multiplying ${\bm{\Phi_C^{\Delta}}}$ by $\Parkw$ gives:
 \begin{gather}\label{Eq:vCDeltaPartC}
\Parkw {\bm{\Phi_C^{\Delta}}} =  \Parkw \left( \ParkInvw \mDeltadqz \LCdot  \ParkInvTw \iSigmadqz  \right) =  
\bm{M_{\Phi_C}^{\Delta}}
\begin{bmatrix}
  \iSigmad & 
  \iSigmaq & 
  \iSigmaz   
\end{bmatrix}^\top 
  \end{gather}
  \begin{gather} \label{Eq:M_phiC_delta}
\bm{M_{\Phi_C}^{\Delta}}  =  
\frac{1}{2} 
\left[
\begin{array}{cc:c}
     {  \mDeltad + 2\mDeltaz \cos(3\omega t)}  
   & {- \mDeltaq - 2\mDeltaz \sin(3\omega t)} 
   & {  \mDeltad} 
     \\ 
     {- \mDeltaq + 2\mDeltaz \sin(3\omega t)}  
   & {- \mDeltad + 2\mDeltaz \cos(3\omega t)}
   & {  \mDeltaq} 
     \\  \hdashline 
     {\mDeltad \cos(3\omega t) + \mDeltaq \sin(3\omega t)}   
   & {\mDeltaq \cos(3\omega t) - \mDeltad \sin(3\omega t)}    
   & {\mDeltaz} 
   \end{array}
   \right]
 \end{gather}
\noindent where $\bm{M_{\Phi_C}^{\Delta}}$ is expressed in \eqref{Eq:M_phiC_delta}. Here, $\bm{M_{\Phi_C}^{\Delta}}$ requires further mathematical manipulation to achieve the desired SSTI performance, as the $3\omega$ signals also appear. Moreover, they affect not only the zero-sequence as in the previous case, yet the $dq$ components as well.
Replacing the definitions given in \eqref{Eq:vCDeltaPartA}, \eqref{Eq:vCDeltaPartB} and  \eqref{Eq:vCDeltaPartC} in $\ParkInvw{\bm{\Phi_A^{\Delta}}} = \ParkInvw{\bm{\Phi_B^{\Delta}}} + \ParkInvw{\bm{\Phi_C^{\Delta}}}$ and solving for the voltage difference dynamics in their $dqz$ coordinates results in \eqref{Eq:vCDeltaALL}:
\begin{equation} \label{Eq:vCDeltaALL}
{\frac{d  \vCDeltadqz}{dt}} = 
\frac{1}{2\Carm}
\left(
\bm{M_{\Phi_B}^{\Delta}} 
{\begin{bmatrix} 
\iDeltad &
\iDeltaq & 
\color{gray}{\iDeltaz}   
\end{bmatrix}^\top} \!
+ \bm{M_{\Phi_C}^{\Delta}}
{\begin{bmatrix} 
  \iSigmad & 
  \iSigmaq & 
  \iSigmaz   
\end{bmatrix}^\top} 
\right)
-\Jw \vCDeltadqz 
\end{equation}
\par Since neither $\bm{M_{\Phi_B}^{\Delta}}$ or $\bm{M_{\Phi_C}^{\Delta}}$ are SSTI, equation \eqref{Eq:vCDeltaALL} is not directly providing a SSTI solution. This issue is treated in the remainder of this section.

\subsubsection{Deriving the SSTI dq dynamics of \eqref{Eq:vCDeltaALL} } 
\par First, the $dq$ dynamics of \eqref{Eq:vCDeltaALL} are addressed. As discussed earlier, since it is assumed that $\iDeltaz = 0$, only $\bm{M_{\Phi_C}^{\Delta}}$ is hindering a SSTI representation for the $dq$ dynamics due to the appearance of the $\cos(3\omega t)$ and $\sin(3\omega t)$ in the $2 \times 2$ sub-matrix at the upper left corner of $\bm{M_{\Phi_C}^{\Delta}}$ in \eqref{Eq:M_phiC_delta}, referred to as $\bm{M_{\Phi_C}^{\Delta 2\times 2}}$. One possible solution is to assume that the MMC control will always set $\mDeltaz$ to zero, as was done in \cite{BergnaCOMPEL2016}, as $\mDeltaz$ is multiplying all of the $3\omega$ oscillating signals. However, this lead to a restrictive model from a control perspective, and therefore such assumption is avoided here. Taking inspiration from common engineering practices to increase controllability in VSCs \cite{Houldsworth1984}, the proposed solution is to redefine $\mDeltaz$ as a third harmonic injection, as given in \eqref{Eq:mDeltaz}, where $\mDeltaZd$ and $\mDeltaZq$ are two SSTI variables that will define the amplitude and phase angle of third harmonic oscilllations in $\mDeltaz$.
\begin{equation}\label{Eq:mDeltaz}
\mDeltaz \defeq \mDeltaZd \cos(3\omega t) + \mDeltaZq \sin(3\omega t)
\end{equation}  
\par Replacing the new definition \eqref{Eq:mDeltaz} in \eqref{Eq:M_phiC_delta}, results in the sub-matrix \eqref{Eq:PHIBSIMP}. 
\begin{align} \label{Eq:PHIBSIMP}
\bm{M_{\Phi_C}^{\Delta 2\times 2}} &= \frac{1}{2}
\left[\begin{array}{cc}
+\left({  \mDeltad + \mDeltaZd }\right) & - \left( \mDeltaq + \mDeltaZq \right) \\ 
-\left({\mDeltaq - \mDeltaZq}\right)& -\left(\mDeltad - \mDeltaZd\right) 
\end{array}\right]+ 
&\color{gray}{\underbrace{\left[\begin{array}{cc}
+\cos\left(6\omega t\right) & +\sin\left(6\omega t\right) \\
+\sin\left(6\omega t\right) &  -\cos\left(6\omega t\right)
\end{array} \right]
\left[
\begin{array}{cc}
+\mDeltaZd & +\mDeltaZq\\
+\mDeltaZq & -\mDeltaZd
\end{array}
\right]}_{\approx 0}}
\end{align}
\par Furthermore, the oscillatory signals at $3\omega$ are replaced by signals at $6\omega$, which can be neglected as will be confirmed via time-domain simulations.

\subsubsection{Deriving SSTI expressions for the zero-sequence dynamics of \eqref{Eq:vCDeltaALL}}

\par The zero sequence dynamics equation of \eqref{Eq:vCDeltaALL}, is given again in \eqref{Eq:vCDeltaz} for convenience.
\begin{align}\label{Eq:vCDeltaz}
 \frac{d  \vCDeltaz}{dt} =& \frac{1}{\Carm}\left[\frac{1}{8} \left(  \mSigmad\iDeltad + \mSigmaq\iDeltaq + 2\mDeltad\iSigmad + 2\mDeltaq\iSigmaq   \right)\cos(3\omega t)  + \dots \right. \\
& \left. \dots + {\frac{1}{8} \left( -\mSigmaq\iDeltad + \mSigmad\iDeltaq + 2\mDeltaq\iSigmad - 2\mDeltad\iSigmaq  \right)}\sin(3\omega t) + {\mDeltaz\iSigmaz} \right]\nonumber
\end{align}
\par By replacing the new definition of \eqref{Eq:mDeltaz} into \eqref{Eq:vCDeltaz}, the zero-sequence dynamics of $\vCDeltaz$ can be written as:
\begin{equation}\label{Eq:vCDeltazCosSin}
  \frac{d  \vCDeltaz}{dt} = \frac{1}{\Carm}\left[{\Psi _{d}}\cos(3\omega t) + {\Psi _{q}}\sin(3\omega t)\right]
\end{equation} 
\noindent where $\Psi _{d}$ and $\Psi _{q}$ are defined as below.
\begin{align}
{\Psi _{d}} =&
\frac{1}{8} \left(  +\mSigmad\iDeltad + \mSigmaq\iDeltaq + 2\mDeltad\iSigmad + 2\mDeltaq\iSigmaq  + 4\mDeltaZd\iSigmaz \right)\nonumber \\
{\Psi _{q}} =& 
\frac{1}{8} \left( -\mSigmaq\iDeltad + \mSigmad\iDeltaq + 2\mDeltaq\iSigmad - 2\mDeltad\iSigmaq + 4\mDeltaZq\iSigmaz \right) \nonumber
\end{align}
\par Since the zero sequence dynamics in \eqref{Eq:vCDeltazCosSin} are still time-varying in steady state, further reformulation is necessary to obtain the desired model with SSTI solution. This can be obtained by defining an auxiliary virtual state $v^{\Delta}_{CZ_{\beta}}$, shifted $90\degree$ with respect to the original "single-phase" time-periodic voltage difference signal $\vCDeltaz$  according to the approach from \cite{BergnaCOMPEL2016}. This approach is conceptually similar to the commonly applied strategy of generating a virtual two-phase system for representing single-phase systems in a SRRF \cite{teodorescu2011grid}. However, since the amplitudes of the different sine and cosine components $\Psi _{d}$ and $\Psi _{q}$ are defined by SSTI variables, the virtual signal $v^{\Delta}_{CZ_{\beta}}$ can be identified without any additional delay.
\par The real and virtual voltage difference zero-sequence variables can be labelled as $v^{\Delta}_{CZ_{\alpha}}$ and $v^{\Delta}_{CZ_{\beta}}$, and together they define an orthogonal $\alpha \beta$-system. This $\alpha \beta$-system can be expressed by \eqref{Eq:vCDeltaAlpha}-\eqref{Eq:vCDeltaBeta}, where the first equation is exactly the same as \eqref{Eq:vCDeltazCosSin}, while the second equation replaces the $\cos(3\omega t)$ and $\sin(3\omega t)$ that appear in \eqref{Eq:vCDeltazCosSin} by $\sin(3\omega t)$  and $-\cos(3\omega t)$, respectively. 
\begin{subequations} 
\begin{gather}
  \frac{d  v^{\Delta}_{CZ_{\alpha}}}{dt} =\frac{1}{\Carm} \left[ {\Psi _{d}}\cos(3\omega t) + {\Psi _{q}}\sin(3\omega t) \right]\label{Eq:vCDeltaAlpha}\\
  \frac{d  v^{\Delta}_{CZ_{\beta}}}{dt} = \frac{1}{\Carm} \left[ {\Psi _{d}}\sin(3\omega t) - {\Psi _{q}}\cos(3\omega t) \right]\label{Eq:vCDeltaBeta}
\end{gather} 
\end{subequations}
\par Defining $\bm{v^{\Delta}_{CZ_{\alpha\beta}}} \defeq [v^{\Delta}_{CZ_{\alpha}} v^{\Delta}_{CZ_{\beta}}]^\top$, the equations \eqref{Eq:vCDeltaAlpha} and  \eqref{Eq:vCDeltaBeta} are written in a compact form as shown in \eqref{Eq:vCDeltaAlphaBeta}.
\begin{equation}\label{Eq:vCDeltaAlphaBeta} 
 \frac{d  \bm{v^{\Delta}_{CZ_{\alpha\beta}}}}{dt} = \frac{1}{\Carm }\left\lbrace
\bm{T_{3\omega}}
\begin{bmatrix}
  {{\Psi _{d}}} &
  {{\Psi _{q}}} 
\end{bmatrix}^\top \right\rbrace
\end{equation} 
\noindent where $\bm{T_{3\omega}}$ can be viewed as a Park transformation at $3\omega$ as defined in \eqref{Eq:T3w}.
\begin{equation}\label{Eq:T3w}
\bm{T_{3\omega}} \defeq
\begin{bmatrix}
  {\cos(3\omega t)}  & {\sin(3\omega t)} \\ 
  {\sin(3\omega t)} & {-\cos(3\omega t)} 
\end{bmatrix}
\end{equation} 
\noindent Furthermore, by defining $\bm{v^{\Delta}_{CZ}} \defeq [v^{\Delta}_{CZ_{d}} v^{\Delta}_{CZ_{q}}]^\top$ which verifies:
\begin{equation}\label{Eq:VCdeltaZdZqDef}
\bm{v^{\Delta}_{CZ_{\alpha\beta}}} = \bm{T_{3\omega}^{\minus 1}}\bm{v^{\Delta}_{CZ}},
\end{equation} 
\noindent replacing \eqref{Eq:VCdeltaZdZqDef} into \eqref{Eq:vCDeltaAlphaBeta}, multiplying  by $\bm{T_{3\omega}}$ and solving for the dynamics of $\bm{v^{\Delta}_{CZ}}$ gives:
\begin{gather} \label{Eq:vCDeltaZdZq}
 \frac{d\bm{v^{\Delta}_{CZ}}}{dt}   = \frac{1}{\Carm}\left\lbrace
\begin{bmatrix}
  {{\Psi _{d}}} &
  {{\Psi _{q}}} 
\end{bmatrix}^\top 
-  \Carm {\bm{J_{3\omega}}} \bm{v^{\Delta}_{CZ}}\right\rbrace
\end{gather} 
\noindent where $\bm{J_{3\omega}}$ is defined as in \eqref{Eq:J3w}.
\begin{equation}\label{Eq:J3w}
\bm{J_{3\omega}} \defeq
\begin{bmatrix}
  {0}  & {- 3\omega } \\ 
  {3\omega} & {0} 
\end{bmatrix}.
\end{equation} 
\par Equation \eqref{Eq:vCDeltaZdZq} will produce now a SSTI solution. The original oscillating zero-sequence component $\vCDeltaz$ can always be re-created as a function of $\vCDeltaZd$ and $\vCDeltaZq$ by means of \eqref{Eq:VCdeltaZdZqDef}, as:
\begin{equation}\label{Eq:vCDeltaZcomplete}
\vCDeltaz = \vCDeltaZd {\cos(3\omega t)}  + \vCDeltaZq {\sin(3\omega t)} 
\end{equation}
\subsubsection{Final formulation}
\par It is useful to redefine a new augmented  vector for the SSTI voltage difference states $\vCDeltadqZdZq$ (with capital Z), as:
\begin{gather}\label{Eq:SSTIvCDeltaFinal} 
\vCDeltadqZdZq \defeq  
\begin{bmatrix}
  {\vCDeltad} & 
  {\vCDeltaq} & 
  {\vCDeltaZd } &  
  {\vCDeltaZq }
\end{bmatrix}^{\top}_,
\end{gather} 
\noindent as well as for the ``$\Delta$'' modulation indexes, as:
\begin{gather} 
\mDeltadqZdZq \defeq  
\begin{bmatrix}
	{\mDeltad} & 
	{\mDeltaq} & 
	{\mDeltaZd } &  
	{\mDeltaZq }
\end{bmatrix}^{\top}_.
\end{gather} 
\par With the new definitions $\vCDeltaZd$, $\vCDeltaZq$ and their associated dynamics given \eqref{Eq:vCDeltaZdZq} as well as taking into account the modified (sub-)matrix $\bm{M_{\Phi_C}^{\Delta 2\times 2}}$ given in \eqref{Eq:PHIBSIMP}; the SSTP dynamics of $\vCDeltadqz$ from \eqref{Eq:vCDeltaALL} may be now expressed in their SSTI equivalents, by means of the $4 \times 1$ state vector $\vCDeltadqZdZq$ as shown in \eqref{Eq:vCDeltaALLdqZ}, with $\bm{J_{G}}$ defined in \eqref{Eq:JGilbert}.
\begin{align} \label{Eq:vCDeltaALLdqZ}
  \frac{d  \vCDeltadqZdZq}{dt} =& -  {\bm{J_{G}}} \vCDeltadqZdZq  + \frac{1}{\Carm}\left\lbrace
\frac{1}{8}  
\begin{bmatrix}
\left( \mSigmad + 2\mSigmaz \right) &   - \mSigmaq  \\ 
- \mSigmaq  &   \left( - \mSigmad + 2\mSigmaz \right)  \\ 
+\mSigmad &  \mSigmaq  \\
 -\mSigmaq &  \mSigmad
\end{bmatrix}\iDeltadq
 +...  \right.\\
& \left. \dots+ \frac{1}{4} 
\begin{bmatrix}
+\left({  \mDeltad + \mDeltaZd }\right)& - \left( \mDeltaq + \mDeltaZq \right) &  {\mDeltad} \\ 
-\left({\mDeltaq - \mDeltaZq}\right) & -\left(\mDeltad - \mDeltaZd\right) &	 {\mDeltaq}\\ 
\mDeltad &  \mDeltaq &  2\mDeltaZd  \\
\mDeltaq & - \mDeltad &  2\mDeltaZq
\end{bmatrix}\iSigmadqz \right\rbrace.  \nonumber
\end{align}
\begin{equation}\label{Eq:JGilbert}
{\bm{J_{G}}} \defeq
\begin{bmatrix}
{\bm{J_{\omega}}}&{\bm{0_{2\times 2}}}\\
{\bm{0_{2\times 2}}}&{\bm{J_{3\omega}}}
\end{bmatrix}
\end{equation} 
\subsection{Voltage sum SSTI dynamics derivation}

\subsubsection{Initial formulation}   

\par The SSTI dynamics for the voltage sum can be derived in a similar way as for the voltage difference. The starting point is indeed the SSTP dynamics of the variable given in \eqref{Eq:vCSigmaABCdef} and recalled in \eqref{Eq:vCSigmaABCa} for convenience. The first step consist in expressing the stationary frame $abc$ vectors present in \eqref{Eq:vCSigmaABCa} as functions of their respective $dqz$ equivalents. This is done in \eqref{Eq:vCSigmaABCc}, where $\vCSigmaabc$, $\mDeltaabc$, $\iDeltaabc$, $\mSigmaabc$ and $\iSigmaabc$ have been respectively replaced by $\ParkInvTw \vCSigmadqz$, $\ParkInvw \mDeltadqz$, $\ParkInvw \iDeltadqz$, $\ParkInvTw \mSigmadqz$ and $\ParkInvTw \iSigmadqz$. Notice that here too, the choice of using the inverse Park transformation at $\omega$ ($\ParkInvw$) or at $2\omega$ ($\ParkInvTw$) is according to the frequency separation of the variables given in Table \ref{Table:Omegavs2Omega} and Fig. \ref{Fig:SSModelOverview}.

\par Equation \eqref{Eq:vCSigmaABCc} can be divided in three parts: ${\bm{\Phi_A^{\Sigma}}}$, ${\bm{\Phi_B^{\Sigma}}}$ and ${\bm{\Phi_C^{\Sigma}}}$, as indicated in \eqref{Eq:vCSigmaABCc}. These three parts are treated consecutively in the following.  
\begin{subequations}\label{Eq:vCSigmaABC}
\begin{gather}
2\Carm \frac{d \vCSigmaabc}{dt}   = \mDeltaabc \LCdot \frac{\iDeltaabc}{2}  + \mSigmaabc \LCdot \iSigmaabc \label{Eq:vCSigmaABCa} \\
\underbrace{2\Carm   \frac{d  \ParkInvTw}{dt} \vCSigmadqz  +  2\Carm \ParkInvTw\frac{d  \vCSigmadqz}{dt}}_{\bm{\Phi_A^{\Sigma}}}  = \underbrace{   \ParkInvw \mDeltadqz \LCdot \frac{ \ParkInvw \iDeltadqz }{2} }_{\bm{\Phi_B^{\Sigma}}} +  \underbrace{ \ParkInvTw \mSigmadqz \LCdot  \ParkInvTw \iSigmadqz }_{\bm{\Phi_C^{\Sigma}}} \label{Eq:vCSigmaABCc}
\end{gather}
\end{subequations}
The equation expressed in \eqref{Eq:vCSigmaABCc}, needs to be multiplied by Park's transformation at $-2\omega$, so that it can be solved for $d\vCSigmadqz/dt$. Multiplying ${\bm{\Phi_A^{\Sigma}}}$ by $\ParkTw$ gives \eqref{Eq:vCSigmaPartA}, where $\JTw$ is defined as ${2} \Jw$.
\begin{gather} \label{Eq:vCSigmaPartA}
\ParkTw {\bm{\Phi_A^{\Sigma}}} =  {2\Carm   \JTw \vCSigmadqz  +  2\Carm  \frac{d  \vCSigmadqz}{dt}}
\end{gather}
Furthermore, multiplying ${\bm{\Phi_B^{\Sigma}}}$ by $\ParkTw$ gives \eqref{Eq:vCSigmaPartB}, where $\bm{M_{\Phi_B}^{\Sigma}}$ is expressed in \eqref{Eq:M_phiB_sigma}.
\begin{gather} \label{Eq:vCSigmaPartB}
  	\ParkTw {\bm{\Phi_B^{\Sigma}}}  = \ParkTw \left(  \ParkInvw \mDeltadqz \LCdot \frac{ \ParkInvw \iDeltadqz }{2} \right) =  \bm{M_{\Phi_B}^{\Sigma}} 
  {\begin{bmatrix} 
  \iDeltad &
  \iDeltaq & 
  \color{gray}\iDeltaz   
\end{bmatrix}^\top}   
 \end{gather}
As mentioned earlier, it is assumed for simplicity in this work that there cannot be any zero-sequence grid current; i.e., $\iDeltaz = 0$ (highlighted in gray). 
\begin{gather} \label{Eq:M_phiB_sigma}
\bm{M_{\Phi_B}^{\Sigma}}  =  
\dfrac{1}{4} 
\left[
\begin{array}{cc:c}
{\mDeltad + 2\mDeltaz \cos\left(3\omega t\right)}   & {-\mDeltaq} + 2\mDeltaz \sin\left(3\omega t\right) & \color{gray}{+2\left(\mDeltad \cos(3\omega t) + \mDeltaq \sin(3\omega t)\right)} \\ 
{\mDeltaq + 2\mDeltaz \sin\left(3\omega t\right)}  & {+ \mDeltad - 2\mDeltaz \cos\left(3\omega t \right)}& \color{gray}{-2\left(\mDeltaq \cos(3\omega t) - \mDeltad \sin(3\omega t)\right)} \\ \hdashline
{\mDeltad }   & {\mDeltaq }  & \color{gray}{2\mDeltaz} 
 \end{array}
 \right]
 \end{gather}
\par Equation \eqref{Eq:vCDeltaPartB} does not yet produce a SSTI solution, as the elements in the upper left $2 \times 2$ sub-matrix of $\bm{M_{\Phi_B}^{\Sigma}}$ in \eqref{Eq:M_phiB_sigma}, contain sine and cosine signals oscillating at $3 \omega$. Note that this is also the case for the terms highlighted in gray, but since these are being multiplied by $\iDeltaz=0$, they are not considered in this work.
To overcome this obstacle, the same solution used in the previous section is applied: as all the oscillating terms are being multiplied by $\mDeltaz$, it is convenient to redefine $\mDeltaz$ by a third harmonic injection as in \eqref{Eq:mDeltaz}, as a function of the SSTI virtual variables $\mDeltaZd$ and $\mDeltaZq$. Replacing \eqref{Eq:mDeltaz} into \eqref{Eq:M_phiB_sigma} allows for re-writing \eqref{Eq:vCSigmaPartB} as in \eqref{Eq:vCSigmaPartBsimp}.
\begin{gather} \label{Eq:vCSigmaPartBsimp}
\ParkTw {\bm{\Phi_B^{\Sigma}}}  =   \frac{1}{4}  
\begin{bmatrix} 
  +\left( \mDeltad + \mDeltaZd \right)\iDeltad  - \left( \mDeltaq - \mDeltaZq \right)\iDeltaq \\ 
  -\left( \mDeltaq + \mDeltaZq \right)\iDeltad - \left(\mDeltad - \mDeltaZd \right)\iDeltaq \\ 
  \mDeltad \iDeltad +  \mDeltaq \iDeltaq 
\end{bmatrix} \! +\! 
\color{gray}{ \frac{1}{4}
\underbrace{\begin{bmatrix}
{\cos(6\omega t)}   & {-\sin(6\omega t)}  & {0} \\ 
{-\sin(6\omega t)}  & {-\cos(6\omega t)} & {0} \\ 
{0}                 & {0}                & {0} 
\end{bmatrix}
\begin{bmatrix}
   + \mDeltaZd\iDeltad - \mDeltaZq\iDeltaq \\ 
  - \mDeltaZq\iDeltad - \mDeltaZd\iDeltaq \\ 
    0   
\end{bmatrix}}_{\approx 0}}
\end{gather}
\par Equation \eqref{Eq:vCSigmaPartBsimp} will become time-invariant only if it is assumed that the oscillatory signals at $6\omega$ can be neglected, which has been confirmed via time-domain simulations.
\par In a similar fashion, ${\bm{\Phi_C^{\Sigma}}}$; i.e., the second component on the right side of \eqref{Eq:vCSigmaABCc}, is multiplied by $\ParkTw$, resulting in \eqref{Eq:vCSigmaPartCsimp}, which can be considered SSTI if the sixth harmonic components are neglected. Here again, the validity of the approximation was confirmed via time-domain simulations.
\begin{gather}\label{Eq:vCSigmaPartCsimp}
\ParkTw {\bm{\Phi_C^{\Sigma}}}  =   
 \frac{1}{2}
\begin{bmatrix}
  2\mSigmaz\iSigmad + 2\mSigmad\iSigmaz \\ 
  2\mSigmaz\iSigmaq + 2\mSigmaq\iSigmaz \\ 
  \mSigmad\iSigmad + \mSigmaq\iSigmaq + 2\mSigmaz\iSigmaz   
\end{bmatrix} \! +\!  
\color{gray}{\frac{1}{2}\underbrace{\begin{bmatrix}
\left(\mSigmad \iSigmad-\mSigmaq \iSigmaq \right)\cos(6\omega t)+\left(\mSigmaq \iSigmad + \mSigmad \iSigmaq \right)\sin(6\omega t) \\
\left(\mSigmad \iSigmad-\mSigmaq \iSigmaq \right)\sin(6\omega t)-\left(\mSigmaq \iSigmad + \mSigmad \iSigmaq \right)\cos(6\omega t)\\
0
\end{bmatrix}}_{\approx 0}}
 \end{gather}
\subsubsection{Final formulation }
\par The SSTI dynamics of the voltage sum vector $\vCSigmadqz$ are found by replacing the SSTI equations \eqref{Eq:vCSigmaPartA}, \eqref{Eq:vCSigmaPartBsimp} and  \eqref{Eq:vCSigmaPartCsimp} in \eqref{Eq:vCSigmaABCc} and solving for $d\vCSigmadqz/dt$, resulting in \eqref{Eq:vCSigmaALL}.
\begin{gather} 
    \frac{d  \vCSigmadqz}{dt} = -      \JTw \vCSigmadqz \label{Eq:vCSigmaALL} + ...\\
...+ \frac{1}{\Carm} \left\lbrace\frac{1}{4}
\begin{bmatrix}
	2\mSigmaz&0&  2\mSigmad \\ 
	0&2\mSigmaz &  2\mSigmaq \\ 
	\mSigmad &  \mSigmaq &  2\mSigmaz
\end{bmatrix}\iSigmadqz
+ 
\frac{1}{8} 
\begin{bmatrix} 
  +\left( \mDeltad + \mDeltaZd \right)&  - \left( \mDeltaq - \mDeltaZq \right)\\ 
  -\left( \mDeltaq + \mDeltaZq \right)& - \left(\mDeltad - \mDeltaZd \right) \\ 
  \mDeltad &  \mDeltaq
\end{bmatrix} \iDeltadq \right\rbrace \nonumber
\end{gather}
\subsection{Circulating current SSTI dynamics derivation}
\par The SSTI dynamics for the circulating current are derived in the following. First, the equation of the dynamics in stationary frame given in \eqref{Eq:iSigmaABC} and recalled in \eqref {Eq:iSigmaDyna}, is rewritten by expressing the $abc$ vectors in the equation as a function of their $dqz$ equivalents, as indicated in \eqref{Eq:iSigmaDynb}.
\begin{subequations}
\begin{gather}
\Larm \frac{d\iSigmaabc}{dt} =  \frac{\bm{v_{dc}}}{2} - \vMSigmaabc - \Rarm \iSigmaabc \label{Eq:iSigmaDyna} \\
\Larm \frac{d\ParkInvTw}{dt}\iSigmadqz +  \Larm \ParkInvTw\frac{d\iSigmadqz}{dt}=  
\frac{\bm{v_{dc}}}{2} - \ParkInvTw\vMSigmadqz - \Rarm \ParkInvTw\iSigmadqz \label{Eq:iSigmaDynb}
\end{gather}
\end{subequations}
\par By further multiplying \eqref{Eq:iSigmaDynb} by $\ParkTw$ and solving for $d\iSigmadqz /dt$ gives:
\begin{equation}\label{Eq:iSigmaabcClassic}
\Larm \frac{d \iSigmadqz}{dt} =  
\begin{bmatrix} 
  {0} &
  {0} & 
  \dfrac{\Vdc}{2} 
\end{bmatrix}^\top 
 - \vMSigmadqz - \Rarm \iSigmadqz  - \Larm \JTw \iSigmadqz, 
\end{equation}
\noindent where $\vMSigmadqz = \ParkInvTw \vMSigmaabc$, and  $\vMSigmaabc$ is defined in \eqref{Eq:vMSigmaabc}.
Nonetheless, in order to assess if \eqref{Eq:iSigmaabcClassic} is SSTI, $\vMSigmadqz$ needs to be rewritten by expressing the $abc$ vectors in the equation as a function of their $dqz$ equivalents, as indicated in \eqref{Eq:vMSigmadqzSSTIcheck}.
\begin{align}\label{Eq:vMSigmadqzSSTIcheck}
\vMSigmadqz =& \frac{1}{2} \ParkTw \left(\ParkInvTw \mSigmadqz \LCdot \ParkInvTw\vCSigmadqz +\ParkInvw \mDeltadqz \LCdot \ParkInvw \vCDeltadqz \right) \nonumber \\
=& \bm{M_{\Psi_B}^{\Sigma}}
\begin{bmatrix}
\vCSigmad & \vCSigmaq & \vCSigmaz
\end{bmatrix}^\top
 + \bm{M_{\Psi_C}^{\Sigma}}
 \begin{bmatrix}
\vCDeltad & \vCDeltaq & \vCDeltaz
\end{bmatrix}^\top_,
\end{align} 
\noindent where $\bm{M_{\Psi_B}^{\Sigma}}$ and $\bm{M_{\Psi_C}^{\Sigma}}$ are expressed in \eqref{Eq:M_psiB_sigma} and \eqref{Eq:M_psiC_sigma}, respectively.
\begin{gather} \label{Eq:M_psiB_sigma}
\bm{M_{\Psi_B}^{\Sigma}}  =  
\frac{1}{4} 
\left[
\begin{array}{ccc}
2\mSigmaz & 0  & 2\mSigmad \\
0 & 2\mSigmaz  & 2\mSigmaq \\
\mSigmad & \mSigmaq & 2\mSigmaz
 \end{array}
 \right]+\left[
 \color{gray}{\underbrace{\begin{array}{ccc}
 \color{gray}{\mSigmad\cos(6\omega t)+\mSigmaq\sin(6\omega t)}& \color{gray}{-\mSigmaq\cos(6\omega t)+\mSigmad\sin(6\omega t)}&\color{gray}{0}\\
 \color{gray}{-\mSigmaq\cos(6\omega t)+\mSigmad\sin(6\omega t)}&\color{gray}{-\mSigmad\cos(6\omega t)-\mSigmaq\sin(6\omega t)}&\color{gray}{0}\\
 \color{gray}{0}&\color{gray}{0}&\color{gray}{0}
 \end{array}}_{\approx 0}}
 \right]
 \end{gather}
\begin{gather} \label{Eq:M_psiC_sigma}
\bm{M_{\Psi_C}^{\Sigma}}  =  
\frac{1}{4} 
\left[
\begin{array}{ccc}
\mDeltad+2\mDeltaz\cos(3\omega t) & -\mDeltaq + 2\mDeltaz\sin(3\omega t) & 2\mDeltad\cos(3\omega t)+ 2\mDeltaq\sin(3\omega t) \\
\mDeltaq +2\mDeltaz\sin(3 \omega t) & +\mDeltad - 2\mDeltaz\cos(3\omega t) & 2\mDeltad\sin(3\omega t)-2\mDeltaq\cos(3\omega t)\\
\mDeltad&\mDeltaq&2\mDeltaz
 \end{array}
 \right]
 \end{gather}
\par If the sixth harmonic components are neglected, $\bm{M_{\Psi_B}^{\Sigma}}$ given in \eqref{Eq:M_psiB_sigma} can be considered as SSTI. This is confirmed via time-domain simulations. However, this is not the case for $\bm{M_{\Psi_C}^{\Sigma}}$ given in \eqref{Eq:M_psiC_sigma},  as it presents non-negligible third harmonic oscillations. To overcome this obstacle, it is necessary to replace into \eqref{Eq:M_psiB_sigma} and in \eqref{Eq:vMSigmadqzSSTIcheck} the new definitions of both $\mDeltaz$ and $\vCDeltaz$ given in \eqref{Eq:mDeltaz} and \eqref{Eq:vCDeltaZcomplete}, respectively. Doing so, results in the SSTI definition of $\vMSigmadqz$ in \eqref{Eq:vMSigmadqzSSTIcheckII}, where $\bm{M_{\Psi_C}^{\Sigma \star}}$ is given in \eqref{Eq:M_psiC_sigmastar} and is SSTI if the sixth harmonic components are neglected.
 \begin{equation}\label{Eq:vMSigmadqzSSTIcheckII}
\vMSigmadqz = \bm{M_{\Psi_B}^{\Sigma}}
\begin{bmatrix}
\vCSigmad & \vCSigmaq & \vCSigmaz
\end{bmatrix}^\top
 + \bm{M_{\Psi_C}^{\Sigma \star}}
 \begin{bmatrix}
\vCDeltad & \vCDeltaq & \vCDeltaZd & \vCDeltaZq
\end{bmatrix}^\top_,
\end{equation}

\begin{gather} \label{Eq:M_psiC_sigmastar}
\bm{M_{\Psi_C}^{\Sigma\star}}  =  
\frac{1}{4} 
\left[
\begin{array}{cccc}
\mDeltad+\mDeltaZd & \mDeltaZq-\mDeltaq &\mDeltad  &\mDeltaq \\
\mDeltaq+\mDeltaZq & \mDeltad-\mDeltaZd & -\mDeltaq & \mDeltad \\
\mDeltad & \mDeltaq & \mDeltaZd & \mDeltaZq
 \end{array}
 \right]+...\\ 
 \scalemath{0.85}{\left[
\underbrace{\begin{array}{cccc}
\color{gray}{\mDeltaZd\cos(6\omega t)+\mDeltaZq\sin(6\omega t)}& \color{gray}{\mDeltaZd\sin(6\omega t)-\mDeltaZq\cos(6\omega t)}&  \color{gray}{\mDeltad\cos(6 \omega t)+\mDeltaq\sin(6\omega t)}&\color{gray}{-\mDeltaq\cos(6\omega t) + \mDeltad\sin(6\omega t)} \\
\color{gray}{\mDeltaZd\sin(6\omega t)-\mDeltaZq\cos(6\omega t)}&\color{gray}{-\mDeltaZd\cos(6\omega t)-\mDeltaZq\sin(6\omega t)}& \color{gray}{\mDeltad\sin(6\omega t)-\mDeltaq\cos(6\omega t)}&\color{gray}{-\mDeltad\cos(6\omega t)-\mDeltaq\sin(6\omega t)} \\
\color{gray}{0}&\color{gray}{0}&\color{gray}{\mDeltaZd\cos(6\omega t)+\mDeltaZq\sin(6\omega t)} & \color{gray}{\mDeltaZd\sin(6\omega t)-\mDeltaZq\cos(6\omega t)}
 \end{array}}_{\color{gray}{\approx 0}}
 \right]}
 \nonumber
 \end{gather}
 \par Finally, replacing \eqref{Eq:M_psiB_sigma} and \eqref{Eq:M_psiC_sigmastar} in \eqref{Eq:vMSigmadqzSSTIcheckII}, and further in \eqref{Eq:iSigmaabcClassic} gives the SSTI dynamics of the circulating current \eqref{Eq:iSigmaabcSSTI}, provided the sixth harmonic components are neglected.
\begin{gather}\label{Eq:iSigmaabcSSTI}
\frac{d \iSigmadqz}{dt} = \frac{1}{\Larm }\left\lbrace 
\begin{bmatrix} 
  {0} \\ 
  {0} \\ 
  \dfrac{\Vdc}{2} 
\end{bmatrix} 
  - \Rarm \iSigmadqz  - \frac{1}{4} 
\left[
\begin{array}{ccc}
2\mSigmaz  & 0 & 2\mSigmad \\
0 & 2\mSigmaz  & 2\mSigmaq \\
\mSigmad & \mSigmaq & 2\mSigmaz
 \end{array}
 \right]\vCSigmadqz+...\right. \\
 \left. ...-\frac{1}{4} 
\left[
\begin{array}{cccc}
\mDeltad+\mDeltaZd & \mDeltaZq-\mDeltaq &\mDeltad  &\mDeltaq \\
\mDeltaq+\mDeltaZq & \mDeltad-\mDeltaZd & -\mDeltaq & \mDeltad \\
\mDeltad & \mDeltaq & \mDeltaZd & \mDeltaZq
 \end{array}
 \right]\vCDeltadqZdZq \right\rbrace  -  \JTw \iSigmadqz\nonumber
\end{gather} 
\subsection{Grid currents SSTI dynamics derivation}
\par Finally, the derivation of SSTI expressions for the grid current dynamics are presented in the following. The beginning of the proof is the SSTP dynamics equation of the grid current in the stationary reference frame given in \eqref{Eq:iDeltaABC}-\eqref{Eq:vMDeltaabc}, and recalled in \eqref{Eq:iDeltaABCderivationa} for convenience. As for the previous cases, the dynamics are rewritten by expressing the $abc$ vectors present in \eqref{Eq:iDeltaABCderivationa} as a function of their $dqz$ equivalents, as indicated in \eqref{Eq:iDeltaABCderivationb}.
\begin{subequations}
\begin{gather}
\Leqac{\frac{d\iDeltaabc}{dt}} = \vMDeltaabc - \Vgabc  - \Reqac\iDeltaabc,\label{Eq:iDeltaABCderivationa}\\
\Leqac{\frac{d\ParkInvw}{dt}\iDeltadqz}+\Leqac\ParkInvw{\frac{d\iDeltadqz}{dt}} = \ParkInvw\vMDeltadqz -\ParkInvw \Vgdqz - \Reqac\ParkInvw \iDeltadqz \label{Eq:iDeltaABCderivationb}
\end{gather}
\end{subequations}
By further multiplying \eqref{Eq:iDeltaABCderivationb} by $\Parkw$ and solving for $d\iDeltadqz /dt$ gives \eqref{Eq:iDeltadqz}.
\begin{equation}\label{Eq:iDeltadqz}
  \Leqac{\frac{d\iDeltadqz}{dt}} = \vMDeltadqz - \Vgdqz  - \Reqac\iDeltadqz - \Leqac \Jw\iDeltadqz
\end{equation}
\noindent where $\Vgdqz = [\Vgd \; \Vgq \; 0]^\top$, $\vMDeltadqz \defeq \Parkw \vMDeltaabc$ and $\vMDeltaabc$ is defined in \eqref{Eq:vMDeltaabc}. Nonetheless, $\vMDeltadqz$  needs to be assessed in order to verify if \eqref{Eq:iDeltadqz} produces a SSTI solution. For this purpuse, $\vMDeltadqz$ is rewritten by expressing the $abc$ vectors in its definition as a function of their $dqz$ equivalents, as indicated in \eqref{Eq:vMDeltadqz_def}. 
\begin{subequations}
\begin{align}\label{Eq:vMDeltadqz_def}
\vMDeltadqz=& - \Parkw  \frac{1}{2}\left( \ParkInvw \mDeltadqz \LCdot \ParkInvTw\vCSigmadqz + \ParkInvTw\mSigmadqz \LCdot \ParkInvw\vCDeltadqz \right)  \\
\vMDeltadqz =& \bm{M_{\Psi_B}^{\Delta}}
\begin{bmatrix}
\vCSigmad & \vCSigmaq & \vCSigmaz
\end{bmatrix}^\top
+\bm{M_{\Psi_C}^{\Delta}}
\begin{bmatrix}
\vCDeltad & \vCDeltaq & \vCDeltaz
\end{bmatrix}^\top
\end{align} 
\end{subequations}
\noindent where $\bm{M_{\Psi_B}^{\Delta}}$ and $\bm{M_{\Psi_C}^{\Delta}}$ are expressed in \eqref{Eq:M_psiB_delta} and \eqref{Eq:M_psiC_delta}, respectively. Both these matrices present non-negligible third order harmonic components preventing the possibility of considering SSTI solutions from \eqref{Eq:iDeltadqz} . As was done in the previous section, it is necessary to replace into \eqref{Eq:M_psiB_delta} and \eqref{Eq:M_psiC_delta} and \eqref{Eq:vMDeltadqz_def} the new definitions of $\mDeltaz$ and $\vCDeltaz$ given in given in \eqref{Eq:mDeltaz} and \eqref{Eq:vCDeltaZcomplete}, respectively. 
\begin{gather} \label{Eq:M_psiB_delta}
\bm{M_{\Psi_B}^{\Delta}}  =  
\frac{1}{4} 
\left[
\begin{array}{ccc}
-\mDeltad-2\mDeltaz\cos(3\omega t)&-\mDeltaq-2\mDeltaz\sin(3\omega t)&-2\mDeltad \\
+\mDeltaq - 2\mDeltaz\sin(3\omega t)& -\mDeltad+2\mDeltaz\cos(3\omega t)&-2\mDeltaq \\ \hdashline
\color{gray}{-\mDeltad\cos(3\omega t)-\mDeltaq\sin(3\omega t)}&\color{gray}{-\mDeltad\sin(3\omega t)+\mDeltaq\cos(3\omega t)}&\color{gray}{-2\mDeltaz}
 \end{array}
 \right]
 \end{gather}
\begin{gather} \label{Eq:M_psiC_delta}
\bm{M_{\Psi_C}^{\Delta}}  =  
\frac{1}{4} 
\left[
\begin{array}{ccc}
-\mSigmad-2\mSigmaz&-\mSigmaq & -2\mSigmad\cos(3\omega t)-2\mSigmaq\sin(3\omega t) \\
-\mSigmaq & -2\mSigmaz+\mSigmad & +2\mSigmaq\cos(3\omega t)-2\mSigmad\sin(3\omega t)\\ \hdashline
\color{gray}{-\mSigmaq\sin(3\omega t)-\mSigmad\cos(3\omega t)}&\color{gray}{\mSigmaq\cos(3\omega t)-\mSigmad\sin(3\omega t)}& \color{gray}{-2\mSigmaz}
 \end{array}
 \right]
 \end{gather}
By doing so, the (reduced) definition of the modulation voltage $\vMDeltadq$ can be expressed as in \eqref{Eq:vMDeltadqz_defSSTI}, where $\bm{M_{\Psi_B}^{\Delta\star}}$ and $\bm{M_{\Psi_C}^{\Delta\star}}$ are given in \eqref{Eq:M_psiB_deltastar} and \eqref{Eq:M_psiC_deltastar}, and will result in SSTI solutions if the sixth harmonic are neglected.
\begin{equation}\label{Eq:vMDeltadqz_defSSTI}
\vMDeltadq = \bm{M_{\Psi_B}^{\Delta\star}}
\begin{bmatrix}
\vCSigmad & \vCSigmaq & \vCSigmaz
\end{bmatrix}^\top+\bm{M_{\Psi_C}^{\Delta\star}}
\begin{bmatrix}
\vCDeltad & \vCDeltaq & \vCDeltaZd & \vCDeltaZq
\end{bmatrix}^\top
\end{equation}
\begin{align} \label{Eq:M_psiB_deltastar}
\bm{M_{\Psi_B}^{\Delta\star}}  = & 
\frac{1}{4} 
\left[
\begin{array}{ccc}
-\mDeltad -\mDeltaZd&-\mDeltaq-\mDeltaZq&-2\mDeltad\\
\mDeltaq-\mDeltaZq&-\mDeltad+\mDeltaZd&-2\mDeltaq
 \end{array}
 \right]+... \\ \nonumber &...+
 \underbrace{\begin{bmatrix}
 \color{gray}{-\mDeltaZd\cos(6\omega t)-\mDeltaZq\sin(6\omega t)}& \color{gray}{-\mDeltaZd\sin(6\omega t)+\mDeltaZq\cos(6\omega t)}& \color{gray}{0} \\
 \color{gray}{\mDeltaZq\cos(6\omega t)-\mDeltaZd\sin(6\omega t)}& \color{gray}{\mDeltaZq\sin(6\omega t)+\mDeltaZd\cos(6\omega t)}&\color{gray}{0}
 \end{bmatrix}}_{\approx 0}
 \end{align}
\begin{align} \label{Eq:M_psiC_deltastar}
\bm{M_{\Psi_C}^{\Delta\star}}  =&  
\frac{1}{4} 
\left[
\begin{array}{cccc}
-\mSigmad-2\mSigmaz & -\mSigmaq & -\mSigmad & -\mSigmaq \\
-\mSigmaq & -2\mSigmaz+\mSigmad& \mSigmaq&-\mSigmad
 \end{array}
 \right]+ ... \\ \nonumber
 &...+\underbrace{\begin{bmatrix}
 \color{gray}{0}&\color{gray}{0}&\color{gray}{-\mSigmaq\sin(6 \omega t)-\mSigmad\cos(6\omega t)}&\color{gray}{\mSigmaq\cos(6\omega t)-\mSigmad\sin(6\omega t)}\\
 \color{gray}{0}&\color{gray}{0}&\color{gray}{\mSigmaq\cos(6\omega t)-\mSigmad\sin(6\omega t)}&\color{gray}{\mSigmaq\sin(6\omega t)+\mSigmad\cos(6\omega t)} 
 \end{bmatrix}}_{\approx 0}
 \end{align}
\par Finally, replacing \eqref{Eq:M_psiB_deltastar} and \eqref{Eq:M_psiC_deltastar} in \eqref{Eq:vMDeltadqz_defSSTI} and further in \eqref{Eq:iDeltadqz} gives the SSTI dynamics of the grid current \eqref{Eq:iDeltaSSTIDyn}, provided the sixth harmonic components are neglected.
\begin{gather}\label{Eq:iDeltaSSTIDyn}
{\frac{d\iDeltadq}{dt}} = \frac{1}{\Leqac}\left\lbrace - \Vgdq  - \Reqac\iDeltadq  + 
\left[
\begin{array}{ccc}
-\mDeltad -\mDeltaZd&-\mDeltaq-\mDeltaZq&-2\mDeltad\\
\mDeltaq-\mDeltaZq&-\mDeltad+\mDeltaZd&-2\mDeltaq
 \end{array}
 \right]\vCSigmadqz +... \right. \\
\left. ...+ \left[\begin{array}{cccc}
-\mSigmad-2\mSigmaz & -\mSigmaq & -\mSigmad & -\mSigmaq \\
-\mSigmaq & -2\mSigmaz+\mSigmad& \mSigmaq&-\mSigmad
 \end{array}
 \right]\vCDeltadqZdZq \right\rbrace - \Jw\iDeltadq\nonumber
\end{gather}

\subsection{MMC Model with SSTI Solution Summary}

\par
\begin{figure}[t]
\centering
\includegraphics[width=0.99\columnwidth]{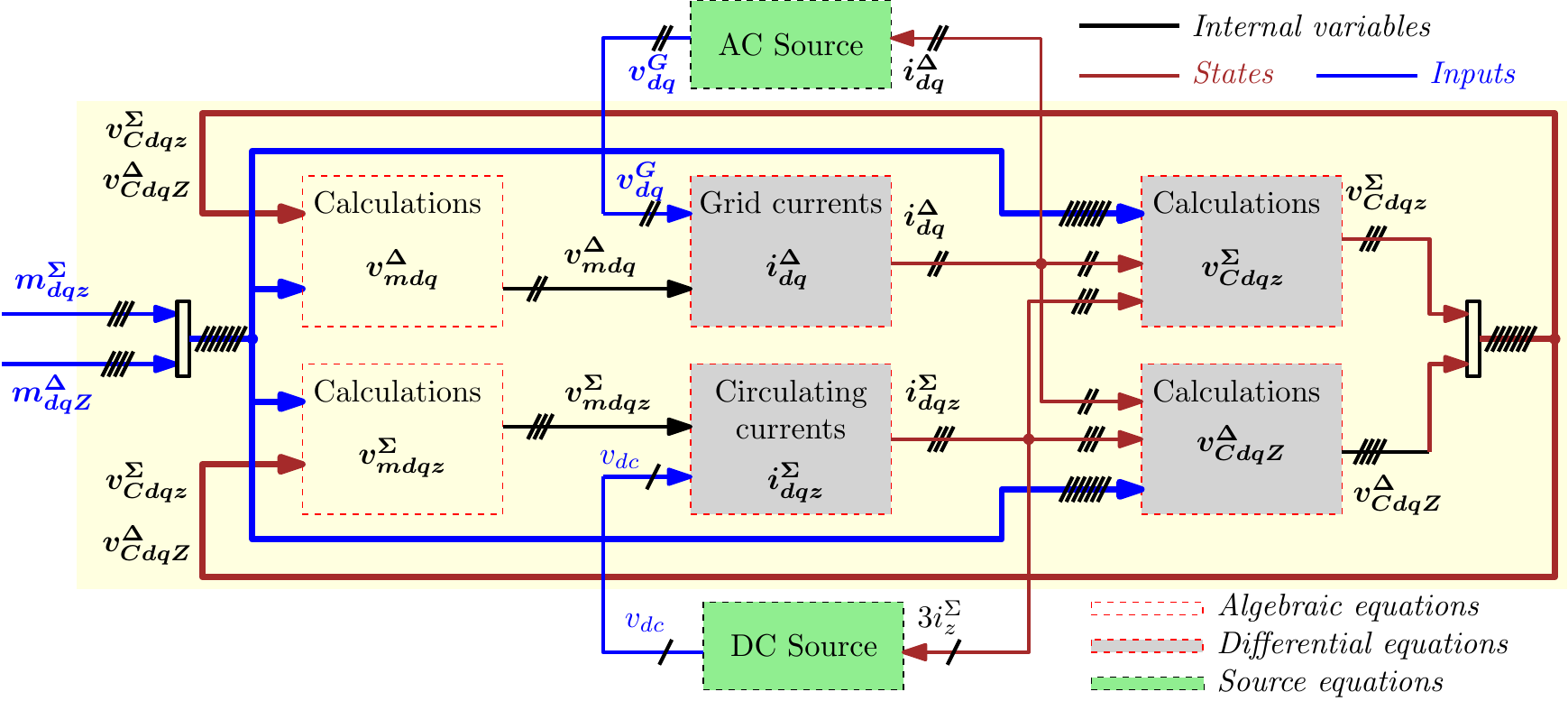}
\caption{Summary of the MMC equations in $dqz$ frame}
\label{Fig:MMCdqBlocks}
\end{figure}

 To summarize, the MMC SSTI dynamics can be represented by means of equations  \eqref{Eq:SSTIvCDeltaFinal}, \eqref{Eq:vCSigmaALL}, \eqref{Eq:iSigmaabcSSTI} and \eqref{Eq:iDeltaSSTIDyn}, corresponding to the 12 SSTI state variables of the arm voltages difference $\vCDeltadqZdZq$, arm voltages sum $\vCSigmadqz$, circulating currents $\iSigmadqz$ and grid currents $\iDeltadq$. Moreover, this model accepts 7 SSTI control inputs represented by the sum and difference of the modulation indices $\mSigmadqz$ and $\mDeltadqZdZq$. In addition, the model receives 3 physical SSTI inputs represented by the voltage at the dc terminals $v_{dc}$ and the $dq$ components of the grid voltage, $\Vgdq$. Finally, the proposed  MMC model with SSTI solution is graphically represented in Fig.~\ref{Fig:MMCdqBlocks}.

\section{Model validation by time-domain simulation}\label{Sec:Results}

\par To validate the developed modelling approach, results from time-domain simulation of the following three different models will be shown and discussed in this section.

\begin{enumerate}
\item The proposed time-invariant MMC model derived in section \ref{Sec:Chapter_MMC_dqz} and represented by equations \eqref{Eq:SSTIvCDeltaFinal}, \eqref{Eq:vCSigmaALL}, \eqref{Eq:iSigmaabcSSTI} and \eqref{Eq:iDeltaSSTIDyn}, corresponding to the SSTI dynamics of the arm voltages difference, arm voltages sum, circulating currents and grid currents. Simulations result obtained with this model are identified in the legend by a $\star$ symbol as a superscript for each variable.

\item The AAM of a three-phase MMC, where each arm is represented by a controlled voltage source and where the internal arm voltage dynamics is represented by an equivalent arm capacitance as indicated in the lower right part of Fig. \eqref{Fig:MMC_Topology_and_AAM_2} \cite{Harnefors2013,Rohner2011,Christe2015}. This model includes non-linear effects except for the switching operations and the dynamics of the sub-module capacitor voltage balancing algorithm, as indicated in Fig. \ref{Fig:StateSpace}. Since this model is well-established for analysis and simulation of MMCs and has been previously verified in comparison to experimental results \cite{Harnefors2013,Rohner2011}, it will be used as a benchmark reference for verifying the validity of the derived model with SSTI solution. The model is simulated in Matlab/Simulink with the SimPowerSystem toolbox. Simulation results obtained with this model are identified in the legend by   ``$AAM$''.

\item The system from Fig.~\ref{Fig:MMC_Topology_and_AAM_2} implemented in EMTP-RV for an MMC with 400 sub-modules per arm, with a capacitance of $0.01302F$ each. The MMC is modeled with the so-called ``Model~\# 2: \emph{Equivalent Circuit-Based Model}'' from \cite{Saad2014}.  This model includes non-linear effects and the switching operations and the dynamics of the sub-module capacitor voltage balancing algorithm from \cite{Tu2012}, as indicated in Fig. \ref{Fig:StateSpace}. Simulation results obtained with this model are identified in the legend by  ``$EMT$''. 
\end{enumerate}

\begin{figure}[t]
	\centering
	\includegraphics[width=0.9\columnwidth]{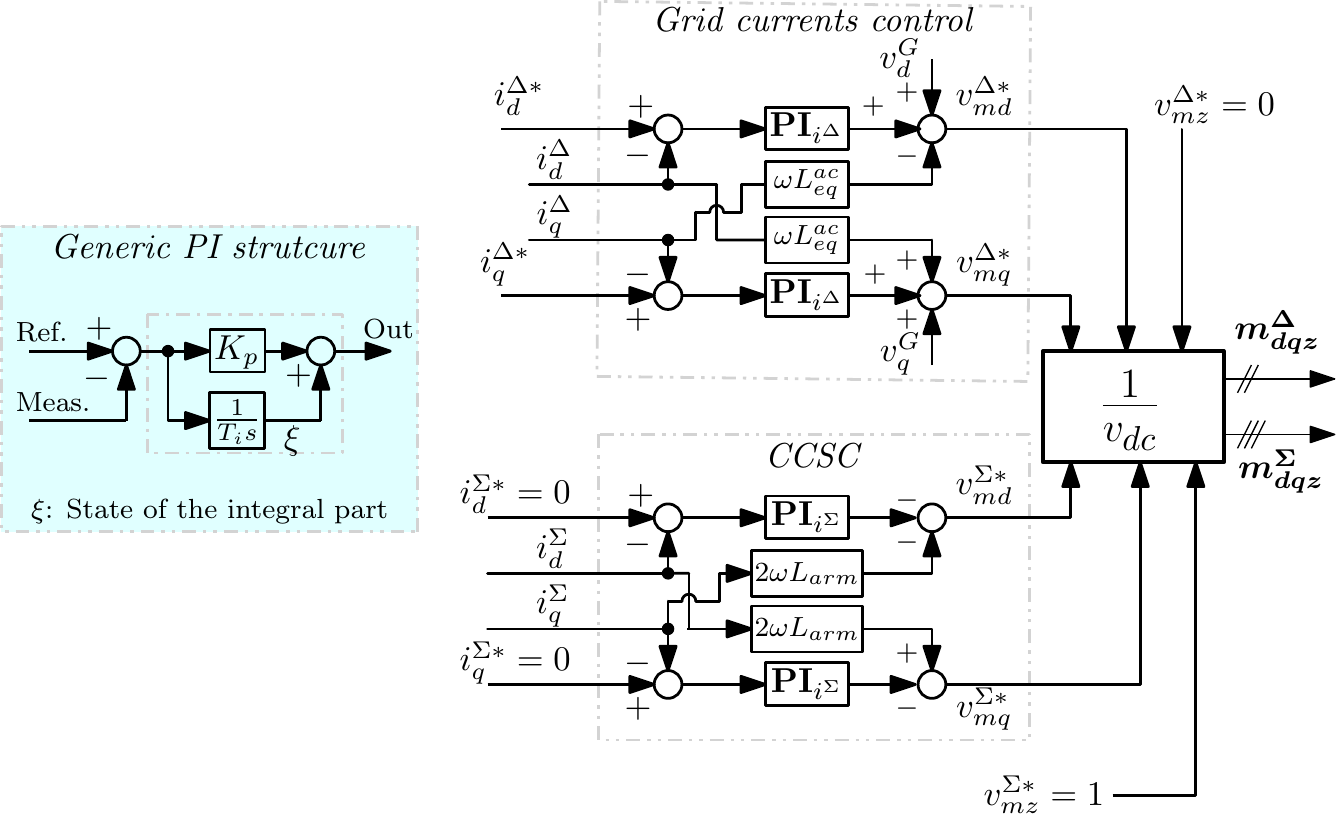}
  \caption{Circulating Current Suppression Control (CCSC) and standard SRRF grid current vector control}
	\label{Fig:CCSC}
\end{figure} %
It is worth mentioning that the verification of the scientific contribution represented by the proposed modelling approach should be done first and foremost with respect to the model it has been derived from; i.e., the AAM. This initial comparison, where the AAM is considered as the reference model, is enough to evaluate the accuracy of the modelling proposal and the simplifications it entails. Thus, the analysis of simulation results that will follow is mainly focused on these two modelling approaches.  Nonetheless, for a more practical-oriented comparison, the detailed switching model has been included as well, to provide an indication to the reader on the accuracy of both the well-established AAM and the proposed modelling approach with respect to a detailed switching model of the MMC.

\par All simulations are based on the MMC HVDC single-terminal configuration shown in Fig. \ref{Fig:MMC_Topology_and_AAM_2}, with the parameters given in Table \ref{Table:MMCParam} under the well known Circulating Current Suppression Control (CCSC) technique described in \cite{Tu2012}, and with standard SRRF vector control for the grid current, similarly to what was presented in \cite{BergnaCOMPEL2016}, and shown in Fig. \ref{Fig:CCSC}. For comparing the models, it should be considered that the reference model is a conventional time-domain simulation model of a three-phase MMC, while the derived model with SSTI solution represents the MMC dynamics by variables transformed into a set of SRRFs. Nonetheless, comparison of transient and steady-state response is simpler when the variables have SSTI representation. Thus, in most cases, the results obtained from the reference model are transformed into the appropriate SRRFs to ease the comparison. However, the results from the models with SSTI solution can also be transformed to the stationary phase coordinates, although this would imply comparison of signals with sinusoidal or multi-frequency oscillations in steady-state. All results are plotted in per unit quantities.

To excite the MMC dynamics in the different models, first the reactive power reference is set from zero to $-0.1$ \textit{p.u.} at $t=0.05s$. Second, the ac-side active power reference is reduced from $1$ \textit{p.u.} to $0.5$ \textit{p.u.} at $t=0.15s$.

\begin{figure}[t]
  \centering
    \begin{subfigure}
    \centering
    \includegraphics[width=0.99\columnwidth]{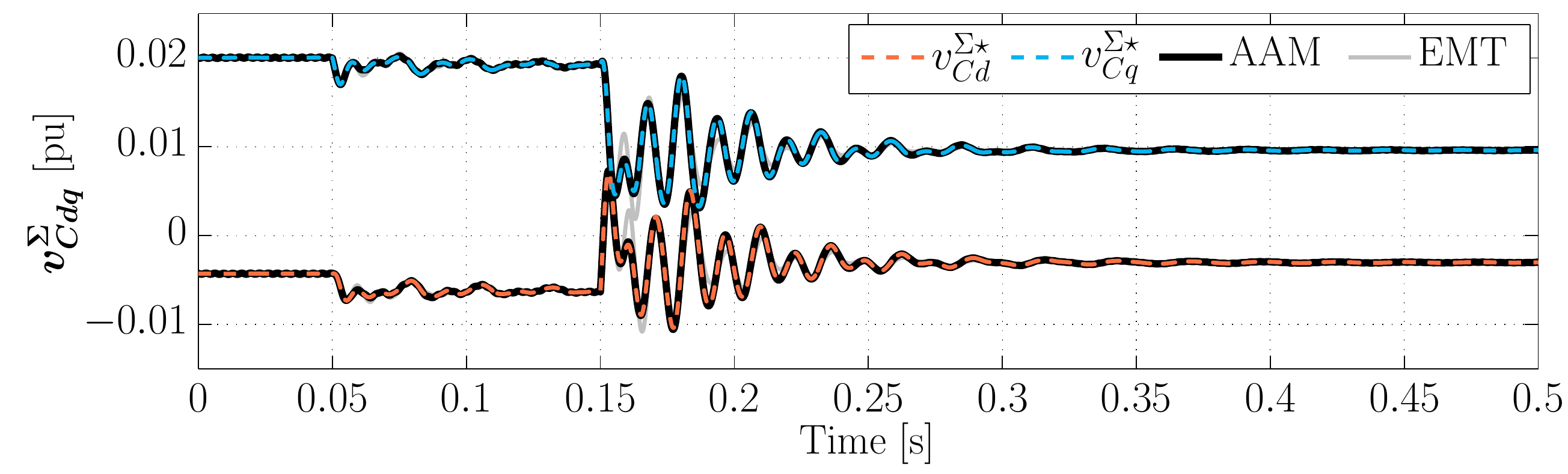}
  \end{subfigure}

  \begin{subfigure}
    \centering
    \includegraphics[width=0.97\columnwidth]{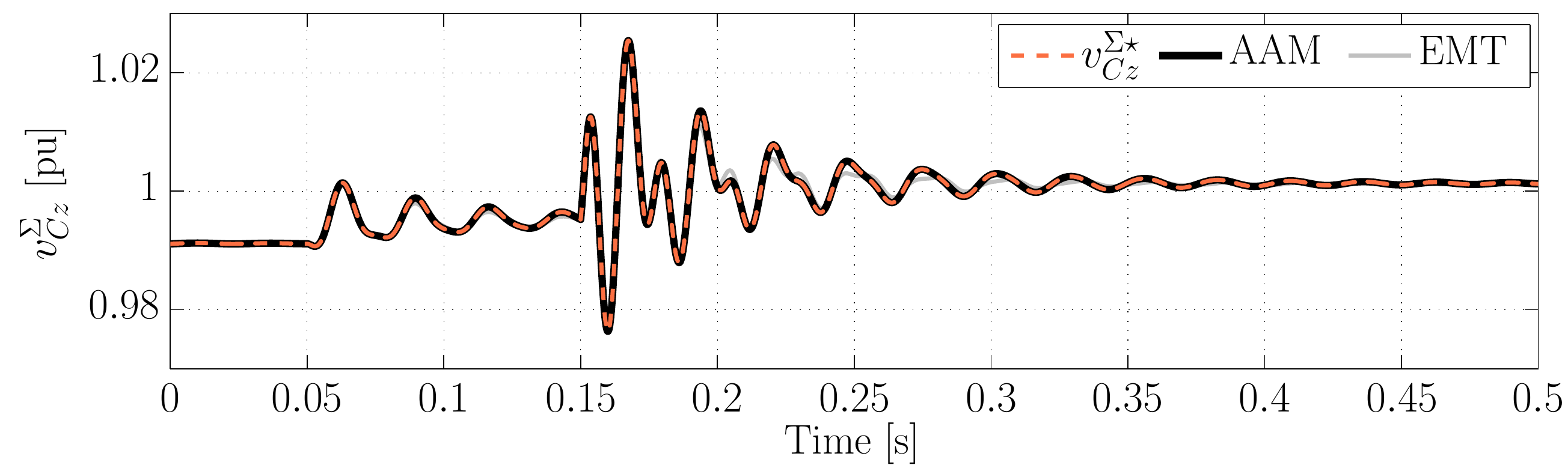}
  \end{subfigure}  
  \caption{Voltage Sum} \label{Fig:vCSigmadqz}
\end{figure}

\par  The dynamics of the voltage sum $\vCSigmadqz$ for the above described case scenario are illustrated in Fig. \ref{Fig:vCSigmadqz}.  More precisely, the $dq$ components of this variable is given in the upper sub-figure while its zero-sequence is shown in the lower one, due to the differences in scale between them. From Fig. \ref{Fig:vCSigmadqz}, it can be seen how the variables calculated with the AAM-MMC used as reference are overlapping those calculated with the model with SSTI solution derived in this paper. This is true for both transient and steady-state conditions. Notice that the steady-state value of $\vCSigmaz$ changes with respect to each of the reference steps, as only the CCSC is implemented assuming no regulation of the capacitive energy stored in the MMC. Furthermore, the non-zero steady-state values of $\vCSigmadq$ reflect the $2\omega$ oscillations that this variable has in the stationary $abc$ reference frame. 


\begin{figure}[t]
  \centering
  \begin{subfigure}
    \centering
    \includegraphics[width=0.98\columnwidth]{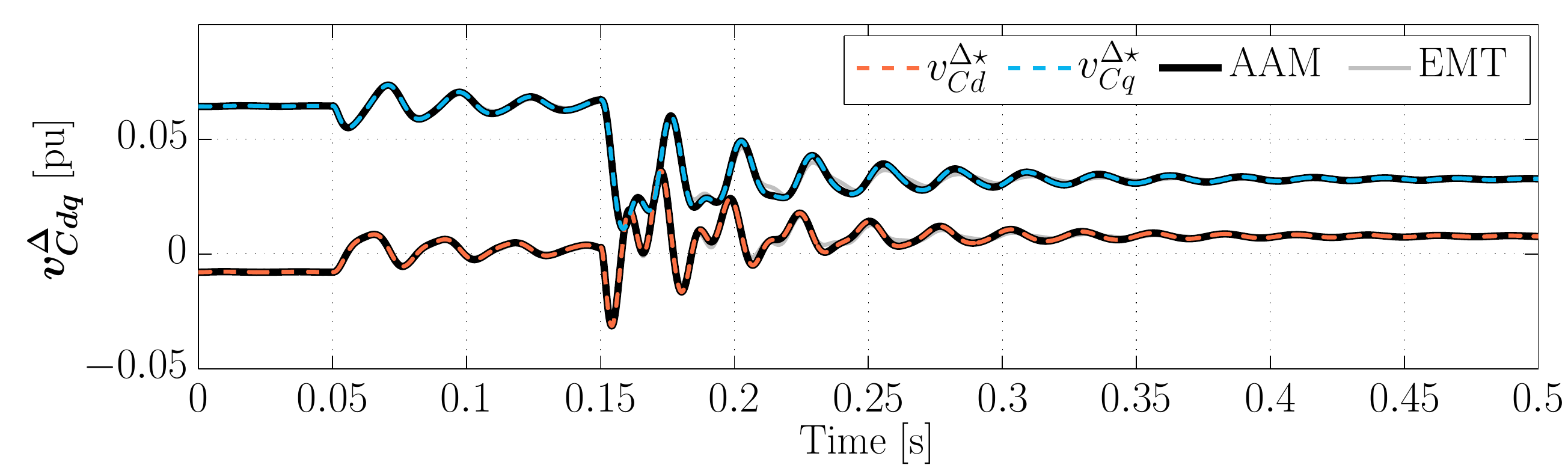}
  \end{subfigure}

  \begin{subfigure}
    \centering
	\includegraphics[width=0.98\columnwidth]{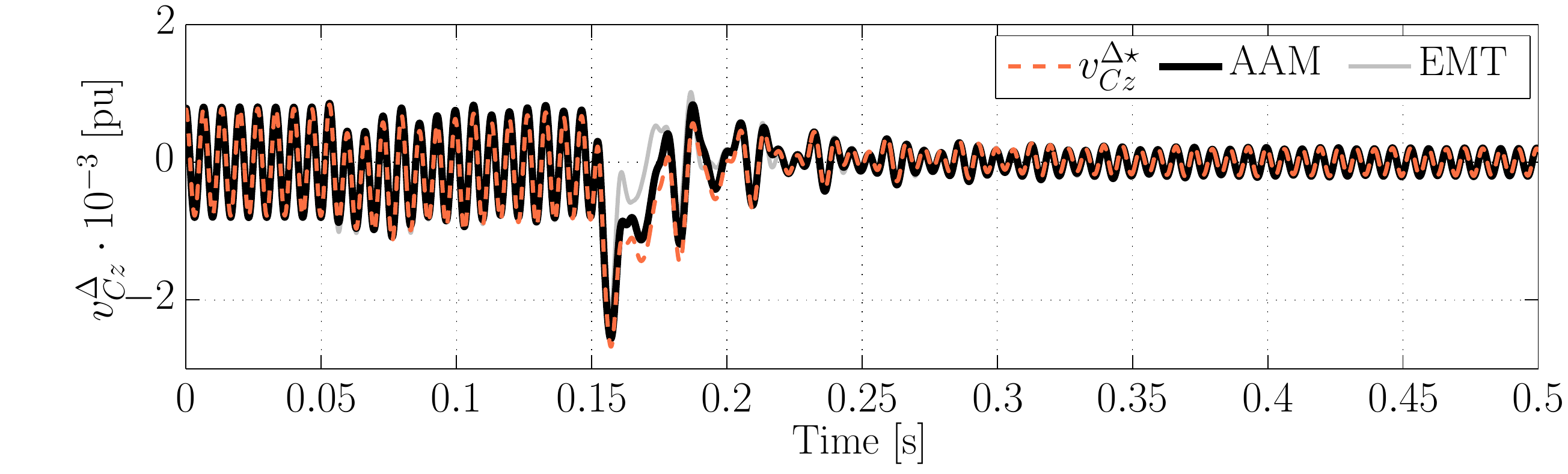}    
  \end{subfigure}  
   \caption{Voltage difference}  \label{Fig:vCDeltadqz}
\end{figure}

\begin{figure}
\centering
\includegraphics[width=0.98\columnwidth]{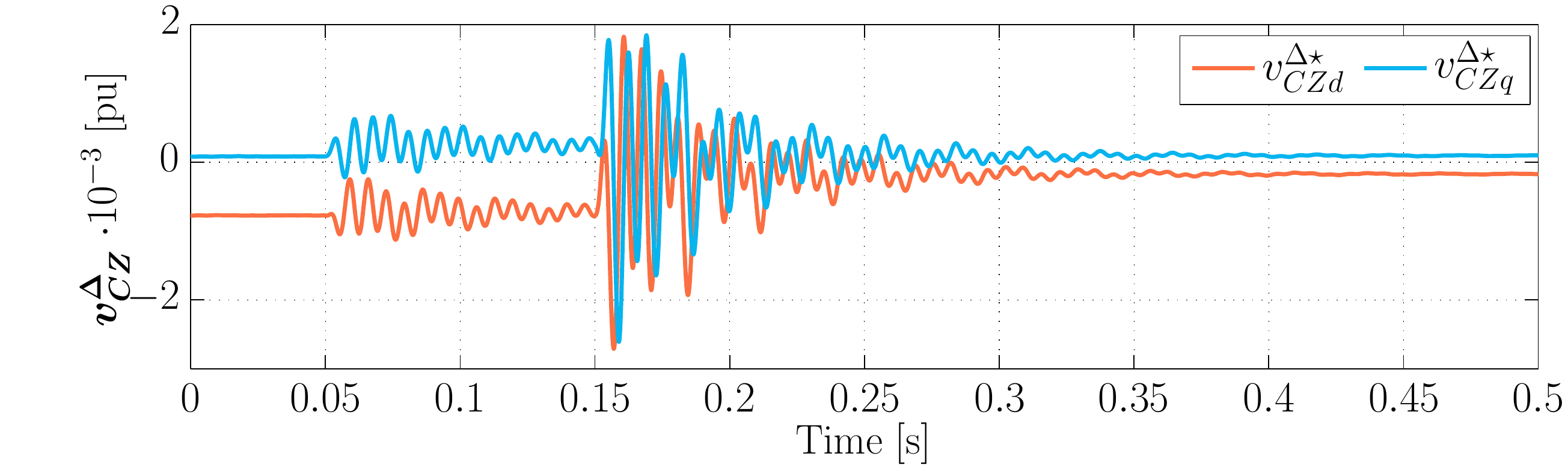}
\caption{SSTI representation of the voltage difference zero sequence}\label{Fig:vCDeltaZdZq}
\end{figure}

\begin{figure}[htb!]
  \centering
  \begin{subfigure}
    \centering
    \includegraphics[width=0.95\columnwidth]{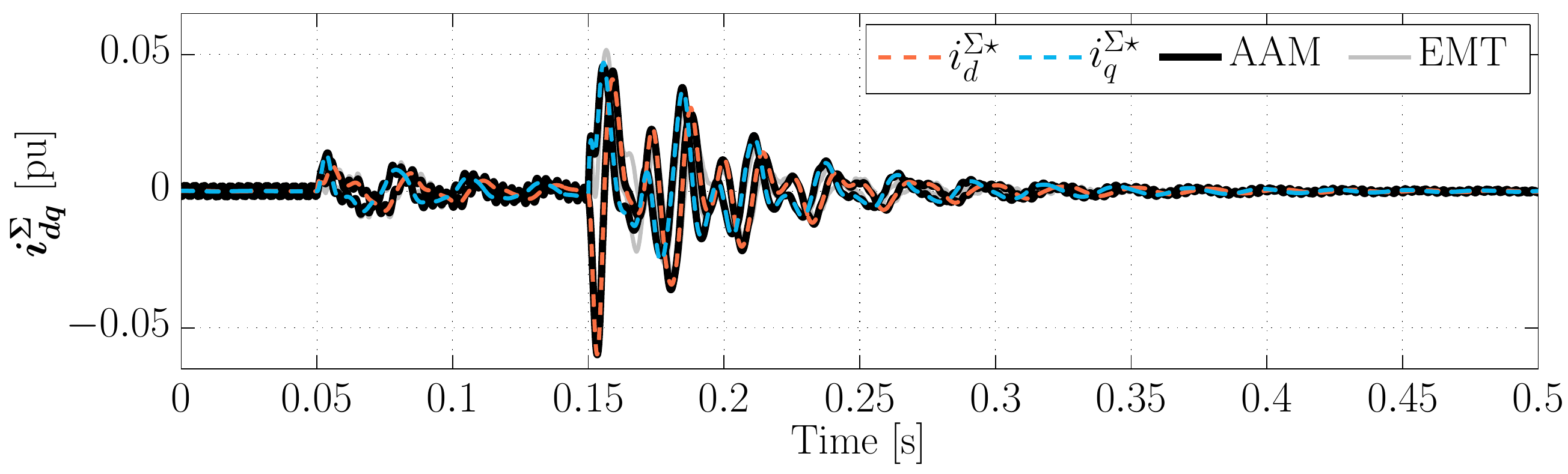}
  \end{subfigure}

  \begin{subfigure}
    \centering
    \includegraphics[width=0.95\columnwidth]{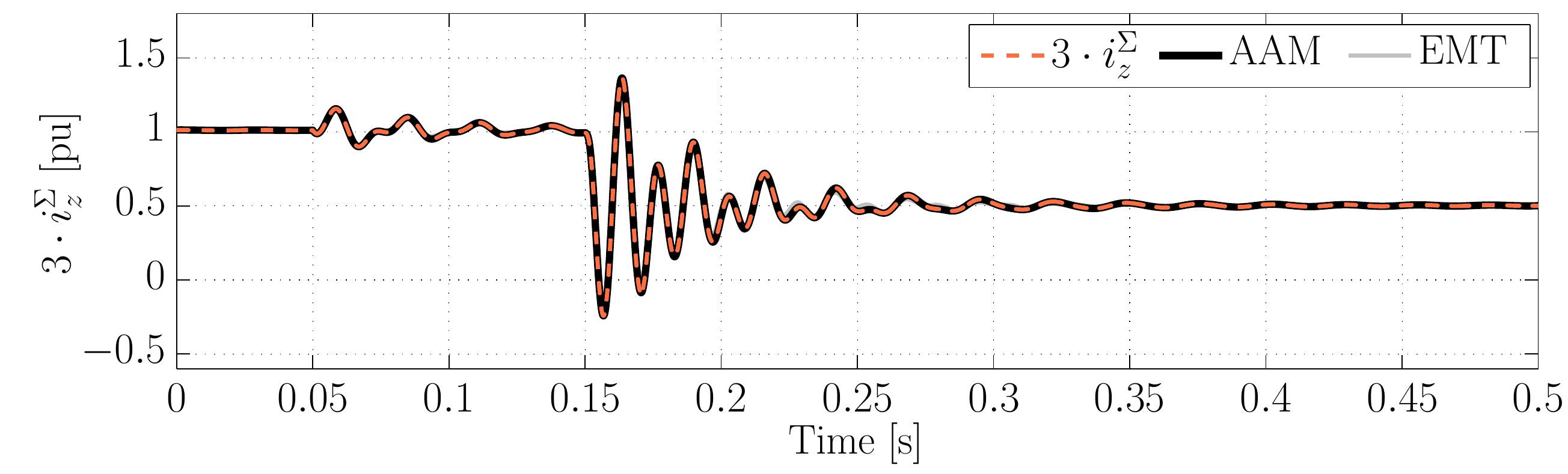}
  \end{subfigure}  
  \caption{Circulating current} \label{Fig:iSigmadqz} 
\end{figure}

\par Similarly, the dynamics of the energy difference $\vCDeltadqz$ are depicted in Fig. \ref{Fig:vCDeltadqz}. More precisely, the upper figure is illustrating the $dq$ components behaviour of this variable under the above described case scenario while the lower figure does the same for the zero-sequence. In terms of accuracy, both of the sub-figures show how the proposed model with SSTI solution accurately captures the behaviour of the AAM-MMC model used as reference. This is particularly true for the case of $\vCDeltadq$ as almost no distinction can be made between the voltage waveforms resulting from the two models. For $\vCDeltaz$ however, it is possible to notice a slight mismatch between the derived model and the AAM, particularly during the transient behaviour between $t=0.15s$ and $t=0.2s$. This is indeed associated to the neglected sixth harmonic components in the mathematical derivation of the proposed model with SSTI solution. Nonetheless, the error is very small and is not having noticeable influence on the general dynamics of the model.

\par Notice that the comparison between the reference and the proposed MMC model with SSTI solution has been done using the SSTP signal $\vCDeltaz$ instead of its equivalent SSTI version $\vCDeltaZdZq$ defined in section \ref{Sec:Chapter_MMC_dqz}. This is done for simplicity, as the dynamics of the virtual system used to create $\vCDeltaZdZq$ do not directly exist in the reference AAM-MMC model. However, for the sake of completeness, the dynamics of the SSTI $\vCDeltaZdZq$ obtained with the proposed model are depicted in Fig. \ref{Fig:vCDeltaZdZq}, where it can be confirmed that both the $\vCDeltaZd$ and $\vCDeltaZq$ sub-variables reach a constant value in steady-state operation.

\par
\begin{figure}[ht]
  \centering
  \begin{subfigure}
    \centering
    \includegraphics[width=0.99\columnwidth]{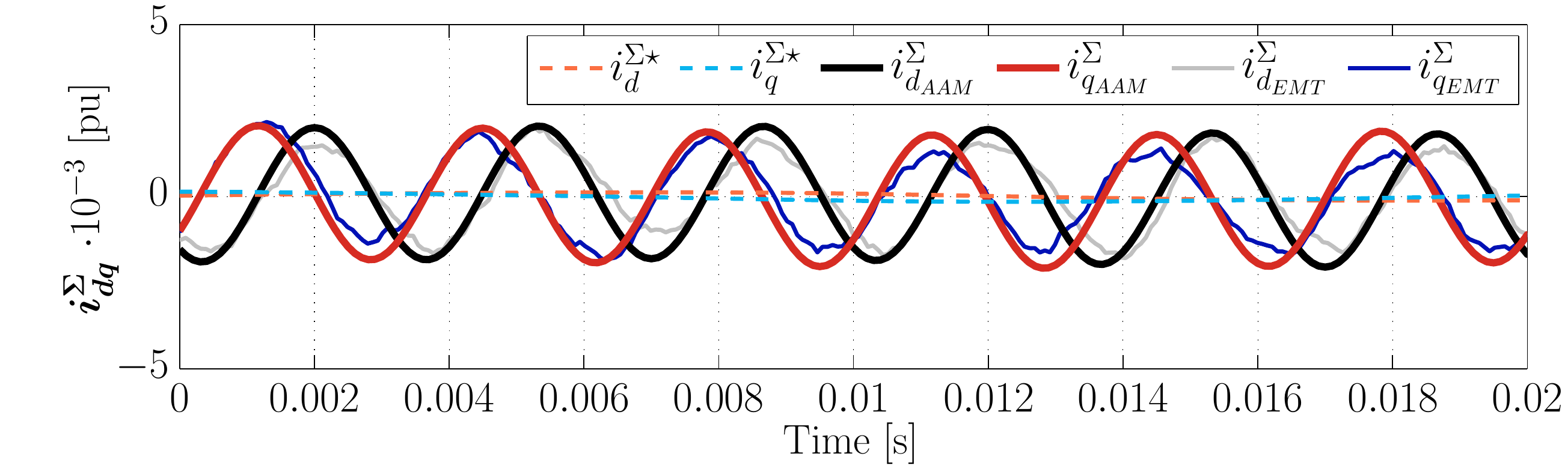}
  \end{subfigure}
  \caption{Circulating current zoom}  \label{Fig:iSigmadqzZoom} 
\end{figure}

 The dynamics of the circulating currents $\iSigmadqz$ are shown in Fig. \ref{Fig:iSigmadqz}, where the upper sub-figure depicts the dynamics of the $dq$ components while the lower figure shows the zero-sequence components multiplied by three, since this signal corresponds to the dc current $i_{dc}$ flowing into the dc terminals of the MMC. From the figure it can be also concluded that the proposed model with SSTI solution replicates quite accurately the dynamic behaviour of the reference model. It can be noticed that the accuracy of the model is very good for the zero-sequence component $\iSigmaz$. However, for the $dq$ component, the 6th order harmonics have been neglected in the modelling. Although these components are very small, they are still present in the reference model, and can be noticed in the figure. Still, the proposed  model captures most of the current dynamics, and is accurately representing the average value of the current components as shown in the zoom of the steady-state operation of $\iSigmadq$ depicted in Fig. \ref{Fig:iSigmadqzZoom}. In this figure, it is possible to see that the reference AAM-MMC model still presents its sixth harmonic components in both the $d$ and $q$ components of the circulating current, whereas the same variables calculated with the proposed  MMC model with SSTI solution only capture the average behaviour. However, given the small value of these oscillations (notice the scale) it can still be considered that the presented modelling approach is sufficiently accurate for most purposes.

\par
\begin{figure}[ht!]
  \centering
  \begin{subfigure}
    \centering
    \includegraphics[width=0.99\columnwidth]{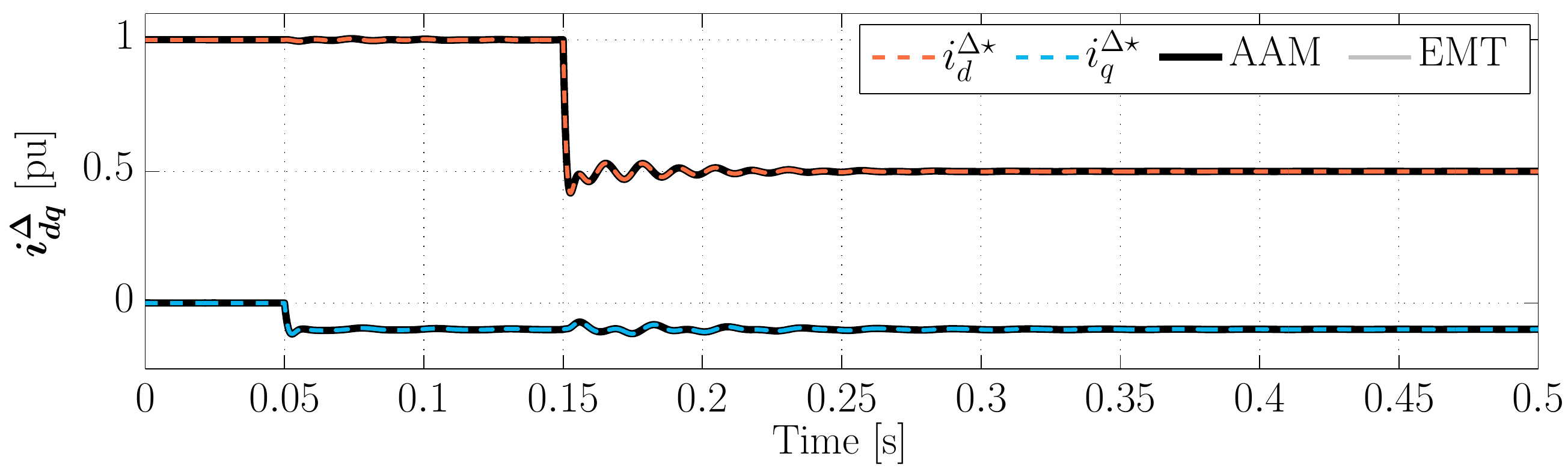}
  \end{subfigure}

  \caption{Grid Current}  \label{Fig:iDeltadq} 
\end{figure}

 Finally, the dynamics of the $dq$ components of the grid current are shown in Fig. \ref{Fig:iDeltadq}. It is possible to see that for this variable the reference model and the proposed  model with SSTI dynamics are practically overlapping.

%
%
%
%
%
%
%
%

\section{Conclusion}\label{Sec:Conclusions}

\par This paper presents a modelling approach for obtaining a state-space representation of an MMC with Steady-State Time-Invariant (SSTI) solution. The presented approach can be considered independent from the modulation and control strategy adopted, as only the physical equations of the MMC have been mathematically manipulated, gaining a more generalized model compared to previous efforts. Results from time-domain simulation of a detailed MMC model with 400 sub-modules per arm are presented as point of reference to illustrate the validity of the derived model. These results demonstrate how the state-space model with SSTI solution accurately captures the MMC internal dynamics while imposing that all state variables settle to a constant equilibrium in steady-state operation. 
This was achieved by a voltage-current $\Sigma$-$\Delta$ formulation which enabled separation of the MMC variables according to their oscillation frequencies as part of the initial model formulation. A procedure for deriving equivalent SSTI $dqz$ representation of all state variables by applying three different Park transformations was presented, referring the variables to three different rotating reference frames, rotating at once, twice and three times the grid fundamental frequency. The resulting model can be suited for detail-oriented studies of MMC control strategies, as it captures the dynamics of the second harmonic circulating currents and the internal energy dynamics of the MMC. 
\par Utilization of the presented model can enable a wide range of studies related to analysis and control system design for MMCs. Since the derived model can be linearised, it can also be utilized for studies of multi-variable control techniques and optimization methods. Furthermore, the model can be utilized for small-signal stability studies by eigenvalue analysis, considering an individual MMC HVDC terminal, or an HVDC terminal integrated in a larger power system configuration. Since the developed MMC model with SSTI dynamics  preserves the  mathematical information about system non-linearities, it is also suited for application of techniques for multi-variable non-linear analysis and control. Moreover, the presented derivations can be useful for developing small-signal models of MMC control systems implemented for operation with individual phase- or arm-quantities.

\appendix \label{Appendix:Parameters}
\par The main parameters are listed in Table \ref{Table:MMCParam}.

\begin{table}[htbp]
  \centering
  \caption{Nominal Values \& Parameters}
    \begin{tabular}{rrrrrr}
    \midrule
    $U_{1n}$  &  $320$[kV]       & $R_f$      & $0.512$[$\Omega$]  & ${\tau}_{i_\Delta}$             & $0.0019$[ms] \\
    $f_n$     &  $50$[Hz]        & $L_f$      & $58.7$[mH]         & ${\tau}_{i_{\Sigma}}$        & $0.0149$[ms]\\
    $N$       &  $400$[-]        & $\Rarm$  & $1.024$[$\Omega$]  & ${k}_{p\Sigma}$      & $0.1253$[pu]\\
    $\Carm$ &  $32.55$[$\mu F$]& $\Larm$  & $48.9$[mH]           & ${k}_{p\Delta}$      & $0.8523$[pu]\\
    \bottomrule
    \end{tabular}%
  \label{Table:MMCParam}%
 \end{table}%

\balance
\bibliography{Biblio_Generalized_MMC_model}
\bibliographystyle{IEEEtran}

\end{document}